\input harvmac
\input psfig
\newcount\figno
\figno=0
\def\fig#1#2#3{
\par\begingroup\parindent=0pt\leftskip=1cm\rightskip=1cm\parindent=0pt
\global\advance\figno by 1
\midinsert
\epsfxsize=#3
\centerline{\epsfbox{#2}}
\vskip 12pt
{\bf Fig. \the\figno:} #1\par
\endinsert\endgroup\par
}
\def\figlabel#1{\xdef#1{\the\figno}}
\def\encadremath#1{\vbox{\hrule\hbox{\vrule\kern8pt\vbox{\kern8pt
\hbox{$\displaystyle #1$}\kern8pt}
\kern8pt\vrule}\hrule}}
\def\underarrow#1{\vbox{\ialign{##\crcr$\hfil\displaystyle
 {#1}\hfil$\crcr\noalign{\kern1pt\nointerlineskip}$\longrightarrow$\crcr}}}
%
\overfullrule=0pt

%
\def\tilde{\widetilde}
\def\bar{\overline}
\def\Z{{\bf Z}}

\def\S{{\bf S}}
\def\R{{\bf R}}

\font\zfont = cmss10 
\font\litfont = cmr6

\def\bigone{\hbox{1\kern -.23em {\rm l}}}
\def\ZZ{\hbox{\zfont Z\kern-.4emZ}}
\def\half{{\litfont {1 \over 2}}}

\Title{hep-th/9810188, IASSNS-HEP-98-82}
{\vbox{\centerline{$D$-BRANES AND K-THEORY}
\bigskip }}
\smallskip
\centerline{Edward Witten}
\smallskip
\centerline{\it School of Natural Sciences, Institute for Advanced Study}
\centerline{\it Olden Lane, Princeton, NJ 08540, USA}\bigskip

\medskip

\noindent
By exploiting recent arguments about stable nonsupersymmetric
$D$-brane states, we argue that $D$-brane charge takes values
in the K-theory of spacetime, as has been suspected before.  
In the process, we gain a 
new understanding of some novel objects proposed recently -- such as
the Type I zerobrane -- and we describe some new objects -- such
as a $-1$-brane in Type I superstring theory.
\Date{October, 1998}

\newsec{Introduction}

One of the most important insights about nonperturbative
behavior of string theory is that $D$-branes carry Ramond-Ramond
charge \ref\polch{J. Polchinski, ``Dirichlet Branes And Ramond-Ramond
Charges,'' Phys. Rev. Lett. {\bf 75} (1995) 4724, hep-th/9510017.}.
Massless Ramond-Ramond fields are differential forms, and therefore
the Ramond-Ramond charges would appear to be cohomology classes -- measured
by integrating the differential forms over suitable cycles in the spacetime
manifold $X$.

\def\K{{\rm K}}
\nref\who{M. Bershadsky, C. Vafa, and V. Sadov, ``D-Branes And
Topological Field Theories,'' hep-th/9511222.}
\nref\odougo{M. R. Douglas, ``Branes Within Branes,'' hep-th/9512077.}
\nref\harvey{M. B. Green, J. A. Harvey, and G. Moore, ``$I$-Brane Inflow
And Anomalous Couplings On D-Branes,'' hep-th/9605033.}
\nref\cheung{Y.-K. Cheung and Z. Yin, ``Anomalies, Branes, and Currents,''
Nucl. Phys. {\bf B517} (1998) 69, hep-th/9710206.}
\nref\moore{G. Moore and R. Minasian, ``K-Theory And Ramond-Ramond Charges,''
hep-th/9710230.}
\nref\mooredouglas{M. R. Douglas and G. Moore, ``$D$-Branes, Quivers,
and ALE Instantons,'' hep-th/9603167.}
There have, however, also been reasons to suspect that one should
 understand $D$-brane charges in terms of the K-theory of spacetime
 rather than cohomology.  First
of all, gauge fields propagate on $D$-brane worldvolumes; this is
more suggestive  of K-theory -- which involves vector bundles
and gauge fields -- than of cohomology.  If, moreover, a $D$-brane
wraps on a submanifold $Y$ of spacetime, then its Ramond-Ramond
charges depend on the geometry of $Y$ and of its normal bundle,
and on the gauge fields on $Y$, in a way that is suggestive of $\K$-theory
\refs{\who - \moore}.  Furthermore, the treatment of
$D$-branes on an orbifold \mooredouglas\ is reminiscent of equivariant 
K-theory.
Finally, one can see Bott periodicity in the brane spectrum     of Type 
IIB, Type IIA, and Type I superstrings.  (For Type II, one has unitary
gauge groups in every even or every odd dimension,  and for Type I,
one flips from $SO$ to $Sp$ and back to $SO$ in adding four to the brane
dimension; these facts are reminiscent of the periodicity formulas
$\pi_i(U(N))=\pi_{i+2}(U(N))$ and $\pi_i(SO(N))=\pi_{i\pm 4}(Sp(N))$.)
Such arguments motivated a proposal \moore\ that $D$-brane charge
takes value in $\K(X)$.  (The proposal was accompanied by a remark that
the torsion in ${\rm KO}(X)$ would provide simple and interesting examples.)
Also, a possible relation of orientifolds with KR-theory was briefly
mentioned in \ref\horavva{P. Horava, ``Equivariant Topological Sigma
Models,'' hep-th/9309124.}.

\nref\senzero{A. Sen, ``Stable Non-BPS States In String Theory,''
JHEP {\bf 6} (1998) 7, hep-th/9803194.}
\nref\senone{A.Sen, ``Stable Non-BPS Bound States Of BPS $D$-branes,''
hep-th/9805019.}
\nref\sentwo{A. Sen, ``Tachyon Condensation On The Brane Antibrane System,''
hep-th/9805170.}
\nref\bergman{O. Bergman and M. R. Gaberdiel, ``Stable Non-BPS $D$-particles,''
hepth/9806155.}
\nref\senthree{A.Sen, ``$SO(32)$ Spinors Of Type I And Other Solitons
On Brane-Antibrane Pair,'' hep-th/9808141.}
\nref\senfour{A. Sen, ``Type I $D$-particle And Its Interactions,''
hep-th/9809111.}
\nref\srednicki{M. Srednicki,  ``IIB Or Not IIB,'' JHEP
{\bf 8} (1998) 5, hep-th/9807138.}
In another line of development, stable but nonsupersymmetric (that is,
non-BPS) states in string theory have been investigated recently
\refs{\senzero - \senfour}.  It has been shown that in many instances
these are naturally understood as bound states of a brane-antibrane system
with tachyon condensation \refs{\senone,\sentwo,\senthree}, and more
concretely as novel stable but nonsupersymmetric $D$-branes \refs{\bergman,
\senfour}.  Brane anti-brane annihilation in the special case of ninebranes
-- which will be important in the present paper -- has been discussed in
\srednicki.

The main purpose of the present paper is to bring these two lines of
development together, by showing that the methods that have been used
in analyzing the brane-antibrane system lead naturally to the identification
of $D$-brane charge as an element of $\K(X)$ -- the K-theory of the spacetime
manifold $X$.  In the process, we will gain some new understanding
of constructions that have been made already, and will propose
some new constructions.

The paper is organized as follows.  Section two is offered
as an appetizer -- some simple questions about Type I superstring theory
are posed, and intuitive answers are given that we will seek to understand
better through the rest of the paper.  The basic relation of the
brane-antibrane system to K-theory is explained in section three.
The identification of $D$-brane charges with K-theory is completed in section
four.  The main idea here is that 
Sen's description of brane-antibrane bound states (as presented most fully
in \senthree) can be identified with a standard construction in K-theory,
involving the Thom isomorphism or Bott class.
In concluding the argument, one also needs a topological condition
that has been noticed previously \ref\witten{E. Witten,
``Baryons And Branes In Anti-de Sitter Space,'' hep-th/9805112.}
and can be understood as a worldsheet global anomaly  
\ref\freedwitten{D. Freed and E. Witten, to appear.} but which
has hitherto seemed rather obscure.  In section five, we generalize
the discussion to orbifolds and orientifolds, and also to include
the Neveu-Schwarz three-form field $H$, which is assumed to vanish in most
of the paper.  
In section six, we discuss worldsheet constructions
for some interesting special cases, including a Type I zerobrane
that has been discussed before, and a new Type I $-1$-brane.

\nref\atiyah{M. F. Atiyah, {\it K-Theory} (W. A. Benjamin, New York, 1967),
reprinted with other useful references in {\it Michael Atiyah Collected Works},
Vol. 2, ``K-Theory'' (Clarendon Press, Oxford, 1988).}
\nref\bott{R. Bott, {\it Lectures On $\K(X)$} (W. A. Benjamin, New York, 
1969).}
For more background on K-theory and fuller explanations of some constructions
that we will meet later, the reader might consult \refs{\atiyah,\bott}.  
I have generally tried to make this paper self-contained and 
readable with no prior familiarity with K-theory, though certain
assertions will be made without proof.

\newsec{Some Questions About Type I Superstrings}

We begin by asking some questions about Type I superstring theory
on $\R^{10}$, and proposing intuitive answers; we will reexamine
these questions in sections four and six.

The gauge group of the Type I superstring is locally isomorphic to $SO(32)$.
The global form of the group is not precisely $SO(32)$ and will
be discussed later.
We will also  compare later with the perturbative $SO(32)$
heterotic string.  Our interest will focus on some of the homotopy
groups of $SO(32)$, namely
\eqn\hazelnut{\eqalign{ \pi_7(SO(32)) & = \Z    \cr
                        \pi_8(SO(32)) & = \Z_2  \cr
                        \pi_9(SO(32)) & = \Z_2. \cr}}
$SO(32)$ bundles on the $i+1$-dimensional sphere $\S^{i+1}$ are
classified by $\pi_i(SO(32))$.   The following
relations of the homotopy groups just introduced 
to index theory will be important
presently: the topological charge of an $SO(32)$ bundle on 
$\S^8$ is measured by the index of the Dirac operator; a nontrivial
$SO(32)$ bundle on $\S^9$ is characterized by having an
odd number of zero modes of the Dirac operator; a nontrivial
bundle on  $\S^{10}$ is similarly characterized by having an odd number
of zero modes of the {\it chiral} Dirac operator.

$SO(32)$ bundles on $\S^{i+1}$
are equivalent to
$SO(32)$ bundles on Euclidean space $\R^{i+1}$ that are trivialized at infinity
(the trivialization means physically that the gauge field is pure gauge
at infinity and the action integral on $\R^{i+1}$ converges). 
So, in ten-dimensional spacetime, we can seemingly use $\pi_7$, $\pi_8$,
and $\pi_9$ to construct strings, particles, and instantons, respectively.
What are these objects?

This question is outside the reach of low energy effective field theory
for the following reason.  Non-zero $\pi_i(SO(32))$ for $i=7,8,9$ leads
to the existence of topologically non-trivial gauge fields on $\R^{i+1}$,
but those objects do not obey the Yang-Mills field equations.  A simple
scaling argument shows that for $n>4$,
the action of any gauge field on $\R^n$ (defined in low energy
effective field theory as ${1\over 4}\int d^nx\,\, \tr F_{ij}F^{ij}$) 
can be reduced by shrinking
it to smaller size.  So the objects associated with $\pi_7, \pi_8$, and 
$\pi_9$,
though they can be constructed topologically using long wavelength gauge
fields, will shrink dynamically to a stringy scale.

Nevertheless, by using low energy effective field theory, we can guess
intuitively the interpretation of these objects:

\bigskip\noindent
{\it The String}

The string associated with $\pi_7(SO(32))$ -- let us call it the
gauge string -- can be identified as follows.\foot{This object has
actually been first constructed in \ref\duffo{
M. J. Duff and J. X. Lu,  ``Strings
From Fivebranes,''
Phys. Rev. Lett. {\bf 66} (1991) 1402; J. A. Dixon, M. J. Duff, and J. C.
Plefka, ``Putting String/Five-Brane Duality To The Test,''
Phys. Rev. Lett. {\bf 69} (1992) 3009.}, where the coupling to
the $B$-field was computed.}
Let $B$ be the two-form field of Type I superstring theory.  
It is a Ramond-Ramond field, and couples to the $D$-string.
However, $B$ also couples to the gauge string
because of the Green-Schwarz anomaly canceling term $\int B\wedge \left(\tr\,
F^4+\dots\right)$, since the gauge string is made from a gauge field on 
$\R^8$ with a nonzero 
integral $\int_{\R^8}\left(\tr\, F^4+\dots\right)$.
In fact, as we will presently calculate, 
the minimal gauge string has $D$-string
charge $\pm 1$.  This strongly suggests that the string
constructed in low energy field theory using a generator of $\pi_7(SO(32))$
shrinks dynamically to an ordinary $D$-string.

\def\ch{{\rm ch}}
To compute the $D$-string charge of the gauge string, let $V$ be an $SO(32)$
bundle on $\R^8$ with a connection of finite action.  Because the connection
is flat at infinity, we can compactify and regard $V$ as an $SO(32)$ bundle
on $\S^8$.  This bundle has $p_1(V)=0$ (since $p_1(V)$ would take
values in $H^4(\S^8)$, which vanishes), and
\eqn\icco{\int_{\S^8}p_2(V)=6k,}
where $k$ is an arbitrary integer. The factor of 6 arises as follows.
As we remarked above, the topological charge of an $SO(32)$ bundle $V$
on $\S^8$ is measured by the Dirac index, which can be
-- depending on the choice
of $V$ -- an arbitrary integer $k$.  On the other hand, using the index
theorem, 
the Dirac index for spinors on $\S^8$  valued in $V$ is
\eqn\nurfo{\int_{\S^8}\ch(V)=\sum_i\int_{\S^8}\left(e^{\lambda_i}+e^{-\lambda_i
}
\right)=-\int_{\bf S^8}{p_2(V)\over 6}.}
Here $\lambda_i$ are the roots of the Chern polynomial, the
Pontryagin classes are $p_1=\sum_i\lambda_i^2$ (which vanishes)
and $p_2=\sum_{i<j}\lambda_i^2\lambda_j^2$, 
and $\ch$ is the Chern character.  So $p_2(V)$ can be any multiple of 6.

On the other hand, the standard anomaly twelve-form (the one-loop anomaly
of the massless
gravitinos and gluinos of the Type I theory) is
\eqn\jkl{-\half\left(p_1(V)-p_1(T)\right)\cdot \left({p_2(V)\over 6}+\dots
\right)}
where the $\dots$ are terms not involving $p_2(V)$.  Since the field
strength $H$ of the $B$-field (normalized so that the periods of $B$ are
multiples of $2\pi$) obeys $dH=\half(p_1(V)-p_1(T))$, the properly
normalized coupling of $B$ to $p_2(V)$ is
\eqn\pokl{\int B\wedge {p_2(V)\over 6}.}
Since $p_2(V)/6$ can be any integer, it follows that the minimal
gauge string has $D$-string charge 1.

\bigskip\noindent
{\it The Particle}

We now consider the particle associated with $\pi_8(SO(32))$; let us
call it the gauge soliton.
We claim that -- in contrast to elementary Type I
string states, which transform
as tensors of $SO(32)$ --
the gauge soliton transforms in a spinorial representation of
$SO(32)$ (a representation of ${\rm Spin}(32)$ in which a $2\pi$ rotation
acts by $-1$).  It can thus, potentially, be compared to the
$D$-particle found in \refs{\senthree,\senfour}, which transforms in this
way.  

To justify the claim, we argue as follows.  The gauge soliton is described
by a nontrivial $SO(32)$ bundle $V$ on $\R^9$, or -- after compactification --
on $\S^9$.  One can pick a connection on $V$ that lives in an $SO(n)$
subgroup of $SO(32)$, for any $n$ with $n\geq 9$.  Such a connection
leaves an  unbroken  subgroup $H=SO(32-n)$. 
We will argue that the gauge soliton transforms in a spinorial representation
of $SO(32)$ by showing that it is odd
under a $2\pi$ rotation in $H$.  We write $V=U\oplus W$,
with $U$ a non-trivial $SO(n)$ bundle and $W$ a trivial $H$ bundle.

As in many such problems involving charge fractionation
\ref\jackiw{R. Jackiw and C. Rebbi, ``Spin From Isospin In A Gauge Theory,''
Phys. Rev. Lett. {\bf 36} (1976) 1116.}, the essence of the matter is to look
at the zero modes of the Dirac operator.  In Type I superstring theory,
the massless fermions that are not neutral under $SO(32)$ are gluinos,
which transform
in the adjoint    representation.  The gluinos that
 transform non-trivially under $H$ and also ``see'' the
$SO(n)$ gauge fields transform as $({\bf n},{\bf 32-n})$ of $SO(n)\times H$
and are sections of $U\otimes W$.  The Dirac operator with values
in $U$ has an odd number of zero modes; generically, this number is one.
So under $H$, one has generically a single vector of fermion zero modes.  Its
quantization gives states transforming in the spinor representation
of $H$, supporting the claim that the gauge soliton is odd under a $2\pi$
rotation in $H$ and hence transforms in a spinorial representation
of $SO(32)$.

This supports the idea that the nonperturbative gauge group of the Type
I superstring is really a spin cover of $SO(32)$.  Duality with the heterotic 
string indicates that  the gauge group is really ${\rm Spin}(32)/\Z_2$
(rather than ${\rm Spin}(32)$).  Possibly this could be seen in the present 
discussion by quantizing the bosonic collective coordinates of the gauge 
instanton (which break $SO(32)$ down to $H$).  We will not attempt to do so.

\bigskip\noindent
{\it The Instanton}

The perturbative symmetry group of the Type I superstring is actually more
nearly 
$O(32)$ than $SO(32)$, as orthogonal transformations of determinant $-1$
are symmetries of the perturbative theory.  (To be more precise, the central
element $-1$ of $O(32)$ acts trivially in Type I perturbation theory,
so the symmetry group in perturbation theory is $O(32)/\Z_2$.)  Duality
with the heterotic string indicates that the transformations of determinant
$-1$ are actually not symmetries, so we must look for a nonperturbative
effect that breaks $O(32)$ to $SO(32)$.  I will now argue that
the instanton associated with $\pi_9(SO(32))$ -- call it the 
gauge instanton (in the present discussion there should be no confusion
with standard Yang-Mills instantons!)
-- has this effect.

The analysis is rather like what we have just seen.  The ten-dimensional
gauge instanton can be deformed to lie in a subgroup $SO(n)$ of $O(32)$ (with
any $n\geq 10$), and so to leave an unbroken subgroup $H=O(32-n)$.
Again we decompose the $O(32)$ bundle as $V=U\oplus W$, with $U$ 
a non-trivial $SO(n)$ bundle and $W$ a trivial $H$ bundle.
To test for invariance under the disconnected component of $O(32)$,
we let $w$ be an element of the disconnected component of $H$ and ask
whether $w$ leaves the quantum measure in the instanton field invariant.
As usual, this amounts to asking whether the measure
for the fermion zero modes is invariant under $w$ --
since everything else is invariant.
The fermions that are not neutral under $O(32)$ are the gluinos.  As in the
discussion of the gauge soliton, the relevant gluinos transform
as $({\bf n},{\bf 32-n})$ under $SO(n)\times H$ and are sections
of $U\otimes W$.  The Dirac equation for (Majorana-Weyl) fermions
with values in $U$ has an odd number of fermion zero modes -- generically
 one.  So the fermion zero modes that are not $H$-invariant
 consist of an odd number of 
 vectors of $H$.  The measure for the zero modes is therefore odd under
 the disconnected component of $H$, supporting the claim that the 
  gauge instanton breaks the invariance under the
 disconnected component of $O(32)$.

The existence of an instanton for Type I seems at first sight to mean
that this theory has a discrete theta angle: one could weight the
instanton amplitude with a $+$ sign or a $-$ sign.  However, the two choices 
give equivalent theories since  a transformation in the disconnected component 
of $O(32)$ changes the sign of the instanton amplitude.

\bigskip\noindent{\it Comparison To The Heterotic String}

Now let us consider how we might interpret the gauge string, soliton,
and instanton for the ${\rm Spin}(32)/\Z_2$
heterotic string.

All three objects are manifest in heterotic string
perturbation theory.  We have interpreted the gauge string
as the Type I $D$-string, which corresponds to the perturbative heterotic
string; the gauge soliton as a particle in the spinor representation,
like some of the particles in the elementary heterotic string spectrum;
and the  gauge instanton as a mechanism that breaks
the disconnected component of $O(32)$ -- a breaking that is manifest
in heterotic string perturbation theory.

So from the point of view of the heterotic string, it seems that
the ostensibly nonperturbative gauge string, soliton, and instanton
can all be continuously converted to ordinary perturbative objects.
But they are relevant to understanding weakly coupled Type I superstring 
theory.

\bigskip\noindent{\it Relation To The Rest Of This Paper}

In all the above, it was not material that the Type I gauge group
is precisely $SO(32)$.  Any orthogonal gauge group 
of large enough rank would have served just as well
(one has $\pi_i(SO(k))=\pi_i(SO(k+1))$ if $k>i$). 
It would have been more convenient
if we could have somehow enlarged the gauge group from $SO(32)$
to $SO(32+n)$ for some $n>9$.  Then we could have carried out the above
arguments with a manifest $SO(32)$ symmetry, instead of seeing only
a subgroup $H$.  The constructions given recently in \refs{\senone,\sentwo,
\senthree} permit one to make
the discussion with  an enlarged gauge group.  This enlargement
is related to K-theory.  
In sections four and six, after learning more,
we will reexamine from new points of view the topological defects that
we have discussed above.

\newsec{Brane-Antibrane Annihilation And K-Theory}

Consider in Type II superstring theory
a $p$-brane and an anti $p$-brane  both wrapped on the same submanifold 
$W$ of a spacetime $X$.  We will use the term $\bar p$-brane as an abbreviation
for anti $p$-brane.
Intuitively, one would expect that as there is no conserved charge
in the system of coincident $p$-brane and $\bar p$-brane, they should
be able to annihilate.  

\nref\green{M. B. Green, ``Point-Like States For Type IIB Superstrings,''
Phys. Lett. {\bf B329} (1994) 435, hep-th/9403040.}
\nref\banks{T. Banks and L. Susskind, ``Brane-Anti-Brane Forces,''
hep-th/9511194.}
\nref\ggut{M. B. Green and M. Gutperle, ``Light-Cone Supersymmetry
And $D$-Branes,'' Nucl. Phys. {\bf B476} 484,
hep-th/9604091.}
\nref\lifs{G. Lifschytz, ``Comparing $D$-Branes To Black Branes,''
Phys. Lett. {\bf B388} (1996) 720, hep-th/9604156.}
\nref\periwal{V. Periwal, ``Antibranes And Crossing Symmetry,''
hep-th/9612215.}
This is supported as follows by the analysis of the brane-antibrane pair.
Lowlying excitations of this system are described, as usual,
by $p$-$p$, $p$-$\bar p$, and $\bar p $-$\bar p$ open strings.
The $p$-$p$ open string spectrum consists of a massless super Maxwell
multiplet plus massive excitations.  The familiar NS sector tachyon is removed
by the GSO projection.  The $\bar p $-$\bar p$ open strings give another
super Maxwell multiplet.  However, for the
$p$-$\bar p$ and $\bar p$-$p$ open strings, one must make the opposite
GSO projection.  Hence, the massless vector multiplet is projected
out, and the tachyon survives \refs{\green - \periwal}. 
It is conjectured  that the
instability associated with the tachyon represents a flow toward
annihilation of the  brane-antibrane pair.  In other words, by giving
the tachyon field a suitable expectation value, one would return to the
vacuum state without this pair.\foot{There is
a puzzle about this process even at a heuristic level \srednicki. 
The gauge group
of the brane-antibrane pair is $U(1)\times U(1)$, with one $U(1)$
on the brane and one on the antibrane. 
The tachyon field $T$ has charges $(1,-1)$, and its
expectation value breaks $U(1)\times U(1)$ to a diagonal $U(1)$ subgroup.
This  $U(1)$ must ultimately be eliminated in the brane-antibrane
annihilation, but it is not clear how this should be described.}  This
has been argued \sentwo\ using techniques in \ref\gava{E. Gava, K. S. Narain,
and M. H. Sarmadi, ``On The Bound States Of $p$- and $(p+2)$-Branes,''
Nucl. Phys. {\bf B504} (1997) 214, hep-th/9704006.}.  Brane-antibrane
annihilation can also \nref\callan{C. G. Callan and J. M. Maldacena, 
``Brane Dynamics From 
The Born-Infeld Action,'' Nucl. Phys. {\bf B513} (1998) 198, hep-th/9708147.}
\nref\savvidy{K. G. Savvidy, ``Brane Death Via Born-Infeld String,''
hep-th/9810163.}  be seen semiclassically \refs{\callan,\savvidy}.

The fact that the $p$-$\bar p$ and $\bar p$-$p$ strings have a reversed
GSO projection can be formalized as follows.  Consider the $p$-$\bar p$ brane
system to have a two-valued Chan-Paton label $i$,  where
$i=1$ for an open string ending on the $p$-brane and $i=2$ for an open
string ending on the $\bar p$-brane.  Thus at the end of the string lives
a charge that takes values in a two-dimensional quantum Hilbert space.
Consider the $i=1$ state to be bosonic and the  $i=2$ state to be fermionic.
Thus, the GSO projection operator $(-1)^F$, which usually acts trivially
on the Chan-Paton factors, acts here
by
\eqn\ikko{(-1)^F=\left(\matrix{ 1 & 0 \cr 0 & -1 \cr}\right).}
The $p$-$p$ and $\bar p $-$\bar p$ open 
strings have diagonal Chan-Paton wave functions.
These wave functions are
 even under $(-1)^F$, leading to the usual GSO projection
on the oscillator modes.  The Chan-Paton wavefunctions for $p$-$\bar p$
and $\bar p $-$ p$ open strings are off-diagonal and 
odd under $(-1)^F$, leading to a reversed
GSO projection for the oscillators.  (Note that it would not matter if we
multiply the right hand side of \ikko\ by an overall factor of $-1$;
in  the action of $(-1)^F$ on string states,
this factor will cancel out, as each open string has two ends.)
Having one bosonic and one fermionic Chan-Paton state would lead, if we
made no GSO projection, to a gauge supergroup $U(1|1)$.  Because of
the GSO projection, the off-diagonal fermionic gauge fields of $U(1|1)$
are absent, and we get instead a structure whose lowest modes correspond
to a ``superconnection'' (in the language of \ref\quillen{D. Quillen,
``Superconnections And The Chern Character,'' {Topology} {\bf 24}
(1985) 89.}), that is to
a matrix of the form
\eqn\kuppo{\left(\matrix{ A & T \cr \bar T & A'\cr}\right),}
where $A$ and $A'$ are the gauge fields 
and $T$ is the $p$-$\bar p$ tachyon.  If $A$ and $A'$ are connections
on bundles $E$ and $F$ ($E$ and $F$ are the bundles 
of  ``bosonic'' and ``fermionic'' Chan-Paton states), then $T$ is
a section of $E\otimes F^*$ and $\bar T$ of $E^*\otimes F$.  ($E^*$ denotes
the dual of a bundle $E$.)
In section six, we will encounter
more exotic actions of $(-1)^F$ on the Chan-Paton wavefunctions.

Now, let us consider a more general case with $n$ $p$-branes and $n$
$\bar p$-branes wrapped on the submanifold $W$ of spacetime.
We allow an arbitrary $U(n)$ gauge bundle $E$ for the $p$-branes,
and (topologically) the {\it same} bundle for the $\bar p$-branes.
The reason for picking the same gauge bundle for both branes and
antibranes is to ensure that the overall system carries no $D $-brane
charges.  (The operator $(-1)^{F_L}$ maps $p$-branes to $\bar p$-branes,
and reverses the sign of all $D$-brane charges, while leaving fixed the
gauge fields on the branes.)  Since this system carries no evident conserved
charges, and there is certainly a tachyon in the $p$-$\bar p$ sector,
one would expect that any such collection of branes can annihilate.
This is the basic technical assumption that we will make in what follows.

Now to proceed, we will specialize first to the case of Type IIB superstrings,
and we will consider first the important special case of what can
be achieved using only 9-branes and $\bar 9$-branes.  $p$-branes with
$p<9$ will be included in the next section.  

We start with an arbitrary configuration with $n$ 9-branes and
the same number of $\bar 9$-branes.  (Tadpole cancellation is the only
reason to require the same number of 9-branes and $\bar 9$-branes.)
In general, the 9-branes carry a $U(n)$ gauge bundle $E$, and the
$\bar 9$-branes carry a $U(n)$ gauge bundle $F$.  We label this
configuration by the pair $(E,F)$.

Now, we ask to what other configurations $(E',F')$ the configuration
$(E,F)$ is equivalent.  The basic equivalence relation we assume
is brane-antibrane creation and annihilation, as described above.
We suppose that any collection of $m$ 9-branes and $m$ $\bar 9$-branes,
with the same $U(m)$ gauge bundle $H$ for both branes and antibranes,
can be created or annihilated.  So the pair $(E,F)$ can be smoothly deformed
to $(E\oplus H,F\oplus H)$.   Since we are only interested in keeping track
of conserved $D$-brane charges -- properties that are invariant under
smooth deformations -- we consider the pair $(E,F)$ to be equivalent
to $(E\oplus H,F\oplus H)$.  

What we have just arrived at is the definition of the K-group
$\K(X)$.  $\K(X)$ is defined by saying that an element of $\K(X)$ is
a pair of complex vector bundles $(E,F)$ over spacetime, subject to
an equivalence relation which is generated by saying that $(E,F)$ is
equivalent to $(E\oplus H,F\oplus H)$ for any $H$.  $\K(X)$ is a group,
the sum of $(E,F)$ and $(E',F')$ being $(E\oplus E',F\oplus F')$.
\foot{It is actually a ring, with the product of $(E,F)$ and $(E',F')$
being $(E\otimes E'\oplus F\otimes F',E\otimes F'\oplus F\otimes E')$,
as if the $E$'s are bosonic and the $F$'s fermionic.  This multiplication
will not be exploited in the present paper.}
One sometimes writes $(E,F)$ as $E-F$.  The subgroup of
$\K(X)$ consisting of elements such that $E$ and $F$ have the same rank
(equal numbers of 9-branes and $\bar 9$-branes) is usually called
$\tilde\K(X)$.  Thus, we conclude that tadpole-cancelling
9-$\bar 9$ configurations, modulo creation and annihilation of brane
pairs, are classified by $\tilde \K(X)$.

At this point, we can explain why and to what extent
it is a good approximation to think
of $D$-brane charge as taking values in cohomology rather than K-theory.
For a vector bundle $E$, let $c(E)=1+c_1(E)+c_2(E)+\dots$ denote
the total Chern class.  The total Chern class of a K-theory class
$x=(E,F)$ is defined as $c(x)=c(E)/c(F)$, the point being that this
is invariant under $(E,F)\to (E\oplus G,F\oplus G)$.  
($1/c(F)=1/(1+c_1(F)+c_2(F)+\dots)$ is defined by expanding it 
in a power series as $1-c_1(F)+c_1(F)^2-c_2(F)+\dots$.)
The component
of $c(x)$ of dimension $2k$ is written $c_k(x)$ and called the $k^{th}$ Chern
class of $x$.  Measuring $D$-brane charge by cohomology
rather than K-theory amounts to measuring a K-theory class by its
Chern classes $c_k(x)$.  This gives a
somewhat imprecise description as there are K-theory
classes whose Chern classes are zero, and is also awkward because
there is no natural description purely in terms of cohomology of precisely
which 
sequences of cohomology classes arise as Chern classes of some $x\in \K(X)$.
However, using cohomology instead of K-theory is an adequate approximation
if one is willing to ignore multiplicative conservation laws (associated
with torsion classes in K-theory) and one in addition does not care
about the precise integrality conditions for $D$-brane charge.

We conclude this section with some technical remarks.
The spacetime $X$ is usually noncompact, for instance $X=\R^4\times Q$ where
$Q$ may be compact.  Because of a finite action or finite energy
restriction, one usually wants objects that are equivalent to the
vacuum at infinity.
Here, ``equivalent to the vacuum
 at infinity'' means that  near infinity, one can relax to the
vacuum by tachyon condensation.  
In many cases, there are no branes in the vacuum, in which case
``equivalent to the vacuum at infinity'' means that in the pair $(E,F)$,
$E$ is isomorphic to $F$ near infinity.  In general the vacuum may contain
branes and thus may be represented by a nonzero K-theory class.
\foot{Tadpole cancellation, or in other words 
the condition that the equations of motion of
Ramond-Ramond fields should have solutions, typically determines the K-theory
class of the vacuum in terms of geometric
data.  For instance, jumping ahead of our story a bit, for Type I
superstrings, the ninebrane charge is 32 for tadpole cancellation,
the fivebrane charge is determined by the equation $dH=\half (\tr F^2-\tr 
R^2)$,
and (if we compactify to two dimensions) the onebrane charge is
determined by the fact that the integrated source  of the 
$B$-field, appearing in the Green-Schwarz coupling 
$\int B\wedge \left(\tr F^4+\dots\right)$, must vanish.}
``Infinity'' means spacetime
infinity if one is considering instantons, spatial infinity in the case
of particles, infinity in the normal directions for strings, and so on.
Requiring that the class $(E,F)$ is equivalent to the vacuum at infinity means
that if we subtract from $(E,F)$ the K-theory class of the vacuum,
we get a K-theory class $(E',F')$ that is trivial at infinity (in the sense
that $E'$ and $F'$ are isomorphic at infinity).  The Ramond-Ramond charge
of 
 an excitation of a given vacuum is best measured by subtracting from
 its K-theory class the K-theory class of the vacuum.

Hence in most physical applications, the Ramond-Ramond charge of
an excitation of the vacuum  is most usefully considered to  take values
not in the ordinary K-group $\K(X)$, but in K-theory with compact support.
More precisely, for instantons one uses
K-theory with compact support, for particles one uses K-theory
with compact support in the spatial directions, etc. 
 A K-theory class with compact support is always represented by a pair
of bundles of equal rank, since bundles that are isomorphic at infinity
must have equal rank.    So the distinction between $\K(X)$
 and $\tilde \K(X)$ is inessential for most physical
applications.  Hence, we will describe our result somewhat loosely by saying
that, up to deformation, 9-brane excitations of a Type IIB spacetime $X$
are classified by $\K(X)$, with the understanding that the precise version of
$\K(X)$  which is relevant depends on the particular situation that
one considers. 

\bigskip\noindent {\it Other String Theories}

\def\KO{{\rm KO}}
Now we consider other theories with $D$-branes, namely Type I and
Type IIA.  

For Type I, the discussion carries over with a few simple changes.
We consider a system with $n$ 9-branes and $m$ $\bar 9$-branes.
Tadpole cancellation now says that $n-m=32$.  The branes support
an $SO(n)$ bundle $E$ and an $SO(m)$ bundle $F$.  By brane-antibrane
creation and  annihilation, we assume that the pair $(E,F)$ is equivalent
to $(E\oplus H,F\oplus H)$ for any $SO(k)$ bundle $H$.

Pairs $E,F$ with this equivalence relation (and disregarding for the
moment the condition $n-m=32$), define the real K-group of spacetime,
written $\KO(X)$.  The subgroup with $n-m=0$ is called $\tilde\KO(X)$.
Any configuration with $n-m=32$ can be naturally mapped  to $\tilde\KO(X)$ by
adding to $F$ a rank 32 trivial bundle.
So pairs $(E,F)$ subject
to the equivalence relation and with $n-m=32$ are classified by $\tilde\KO(X)$.
  As we noted in discussing Type IIB, 
in most physical applications,
one wishes to measure the K-theory class of an excitation relative
to that of the vacuum.  If we do so, then the brane charge of an
excitation is measured by  $\tilde{{\rm
KO}}(X)$
with a compact support condition.  With such a compact support condition
$\KO$ and $\tilde \KO$ are equivalent, so we will describe our result
by saying that $9$-$\bar 9$ configurations of Type I are classified by
$\KO(X)$.\foot{Unlike what we said for Type IIB,
this identification of 9-brane configurations for Type I with KO-theory
does not really require
assumptions about brane-antibrane annihilation, in the following sense.
Since $X$ has dimension 10, the classification of $SO(32)$ bundles on $X$
is governed by  the homotopy groups $\pi_i(SO(32))$ for $i\leq 9$ and the
relations among them.
These homotopy groups are in the ``stable range,'' and one can show
 that $SO(32)$ bundles on $X$ are classified by $\tilde {\rm {KO}}(X)$.}

The discussion for Type IIA is more subtle, because the brane world-volumes
 have odd codimension. I will not attempt a complete description in the
 present paper. The basic idea is to relate branes not to bundles
on $X$ but to bundles on $\S^1\times X$.\foot{It is tempting to believe
that the circle that enters here is related to the circle used in relating
Type IIA to $M$-theory, but I do not know a precise relation.}  
Given a $p$-brane wrapped on an odd-dimensional
submanifold $Z\subset X$, we identify
$Z$ with a submanifold $Z'=w\times Z$ in $\S^1\times X$, where 
$w$ is any point in $\S^1$.  $Z'$ has even codimension in $\S^1\times X$,
and by a construction explained in the next section, a brane wrapped
on $Z'$ determines an element of $\K(\S^1\times X)$.  This element
is trivial when restricted to $X$ (that is, to $w'\times X$ for any
$w'\in \S^1$).  By a more full study of brane-antibrane creation
and annihilation, one expects to show that two Type IIA brane configurations
on $X$ are equivalent if they determine the same element of $\K(\S^1\times X)$.
The subgroup of $\K(\S^1\times X)$ consisting of elements that are trivial
on $X$ is called $\K^1(X)$.\foot{By Bott periodicity, $\K(X)$ and $\K^1(X)$
are the only complex K-groups of $X$. So we don't need more Type II string
theories that would use more K-groups!}  For application to Type IIA,
we must consider the subgroup $\tilde \K^1(X)=\tilde \K(\S^1\times X)$
(since we have no physical interpretation for tenbranes wrapping 
$\S^1\times X$!), and we also want a compact support condition that
generally makes $\tilde \K^1 $ and $\K^1$ equivalent.  Generally, then,
$D$-brane charges of Type IIA are classified by $\K^1(X)$, with an appropriate
compact support condition.

\newsec{Incorporating $p$-Branes With $p<9$}

In the last section, we saw (with certain assumptions about brane
annihilation) that Type IIB configurations of ninebranes,
modulo deformation, are classified by $\K(X)$.  We also explained
the analogs for Type I and Type IIA.
In this section, we will show that the charges are still classified in the
same way if one relaxes the restriction to ninebranes.
The basic  idea is to exploit a construction used by Sen \senthree\
to interpret $p$-branes of $p<9$ as bound states of brane-antibrane
pairs of higher dimension.   In the
discussion, we assume that all spacetime dimensions are much larger than
the string scale.  If this discussion is relaxed, one will meet
new stringy phenomena.  We also assume that the Neveu-Schwarz three-form
field $H$ vanishes (at least topologically); it is incorporated in section 
five.

\subsec{ Review Of Sen's Construction}

We first review Sen's basic construction of a $p$-brane as a bound
state of a $p+2$-brane and a coincident $p+2$-antibrane.  First we
work in $\R^{10}$, without worrying about effects of spacetime topology.

We consider an infinite $p+2$ brane-antibrane pair stretching over an
$\R^{p+3}\subset \R^{10}$.
On the brane-antibrane pair, there is a $U(1)\times U(1)$ gauge field,
with a tachyon field $T$ of charges $(1,-1)$.  We consider a ``vortex''
in which $T$ vanishes on a codimension two subspace $\R^{p+1}\subset
\R^{p+3}$, which will be interpreted as the $p$-brane worldvolume.
We suppose that $T$
approaches its vacuum expectation value at infinity, up to gauge
transformation.  Since $T$ is a complex field, it can have a
``winding number'' around the codimension two locus on which it vanishes,
or equivalently, at infinity.
The basic case is the case that the ``winding number''
is 1.  $T$ breaks $U(1)\times U(1)$ to $U(1)$.  To
keep the energy per unit $p$-brane volume finite, there must be a unit
of magnetic flux in the broken $U(1)$.  Because of this magnetic
flux, the system has a $p$-brane charge of 1, as in \odougo.  
Its $(p+2)$-brane charge
cancels, of course, between the brane and antibrane.
With the  tachyon
close to its vacuum expectation value except close to the core of the
vortex, the system looks like the vacuum everywhere except very near the locus
where $T$ vanishes.  Since this locus carries unit $p$-brane charge, it
seems that a $p$-brane has been 
realized as a configuration of a $(p+2)$-brane-antibrane pair.
 
How would we generalize this to exhibit a $p$-brane as a configuration
of $p+2k$-branes and antibranes for $k>1$?  
One way to do this is to repeat the above construction $k$ times.
We first make a $p$-brane as a bound state of a $(p+2)$-brane and antibrane.
Then, we make the $(p+2)$-brane and antibrane each from a 
$(p+4)$-brane-antibrane
pair.  So at this stage, the $p$-brane is built from two 
$(p+4)$-brane-antibrane
pairs. After $k-2$ more such steps, we get a $p$-brane built 
from $2^{k-1}$ pairs of $(p+2k)$-branes and antibranes.

To exhibit the symmetries more fully and for  applications below, 
it helps to make this construction
``all at once'' and not stepwise.
For this, we consider in general a collection of many
 $(p+2k)$-brane-antibrane pairs, say $n$ such pairs for some
sufficiently large $n$.  The branes carry a $U(n)\times U(n)$ gauge symmetry
under which the tachyon field $T$ transforms as 
$({\bf n},{\bf \bar n})$.  In vacuum, $T$ breaks $U(n)\times U(n)$
down to a diagonal $U(n)$.\foot{The possibility of separating the 
brane-antibrane pairs indicates that the eigenvalues of $T$ are all
equal in vacuum, so that in vacuum $T$ breaks $U(n)\times U(n)$ to
$U(n)$, rather
than a subgroup.  As noted in \srednicki\ and above, there is a puzzle
here, namely how to think about the fate of the diagonal $U(n)$ that
is not broken by $T$.}  The gauge orbit of values of $T$ with minimum
energy is hence a copy of $U(n)$.

To make a $p$-brane, we want $T$ to vanish in codimension $2k$  (on an 
$\R^{p+1}\subset\R^{p+2k+1}$) and to approach
its vacuum orbit at infinity, with a non-zero topological ``twist'' around
the locus on which $T$ vanishes.  Such configurations are classified
topologically
by $\pi_{2k-1}(U(n))$.  According to Bott periodicity, this group equals
$\Z$ for all sufficiently large $n$.  This copy of $\Z$ will label the
possible values of $p$-brane charge.  The value $n=2^{k-1}$
is suggested by the above stepwise construction, and indeed for this value
one can give
a particular simple and -- as we will see -- useful description of the
generator of $\pi_{2k-1}(U(n))$.  Let $S_+$ and $S_-$ be the positive
and negative chirality spinor representations of $SO(2k)$; they are of 
dimension
$2^{k-1}$.  Let $\vec \Gamma=(\Gamma_1,\dots,\Gamma_{2k})$ be the usual
Gamma matrices, regarded as maps from $S_-$ to $S_+$.  If
$\vec x=(x_1,\dots,x_{2k})$ is an element of $\S^{2k-1}$ (that is,
a $2k$-vector with $\vec x^2=1$), then we define the tachyon field
by
\eqn\gurrfo{T(\vec x)=\vec \Gamma\cdot \vec x.}
It has winding number 1 and (according to section 2.13 of
\ref\abs{M. F. Atiyah, R. Bott, and A. Shapiro, {\it Clifford Modules},
Topology {\bf 3} (1964) 3, reprinted in {\it Michael Atiyah Collected
Works}, op. cit.})
generates $\pi_{2k-1}(U(2^{k-1}))$.

That the $p$-brane charge of this configuration is 1 and all higher
(and lower) charges vanish can be verified by using the formulas for
brane charges induced by gauge fields, or in a more elementary way by
verifying that the ``all at once'' construction \gurrfo\ is equivalent
to the stepwise construction that we described first.

Since this configuration
has $p$-brane charge 1 and looks like the vacuum except near
$\vec x=0$, we assume, in the spirit of Sen's constructions, that this
configuration describes a $p$-brane.

\def\V{{\cal V}}
We now wish to place this construction 
in a global context.  The goal is to show that, globally, brane charge
in a spacetime $X$ can always be described by a configuration
of 9-branes and $\bar 9$-branes and so is classified as in section 3.
The technical arguments that follow can be found in \abs;
 physics will intrude
only when we discuss the anomaly.

\subsec{Global Version}

\def\L{{\cal L}}
\def\M{{\cal M}}
We first consider  Sen's original construction (codimension two) in a global
context.
We let $Z$ be a closed submanifold of spacetime, of dimension $q=p+1$,
and we suppose that $Z$ is contained in $Y$, a submanifold of spacetime
of dimension $q+2$. We assume here and in section (4.3) that $Z$ and $Y$
are orientable,
since Type II branes can only wrap on orientable manifolds.
 Then one can define a complex line bundle $\L$ over $Y$,
and a section $s$ of $\L$ that vanishes precisely along $Z$, with a simple
zero.  Moreover, one can put a metric on $\L$ such that, except in a small
neighborhood of $Z$, $s$ has fixed length.

Now, consider a system consisting of a $(p+2)$-brane-antibrane
pair, wrapped on $Y$.  We place on the brane a $U(1)$ gauge field that
is a connection on ${\cal L}$; its $p$-brane charge is that of a $p$-brane
wrapped on $Z$.  We place on the antibrane a trivial
$U(1)$ gauge field, with vanishing $p$-brane charge.  The brane-antibrane
system thus has vanishing $(p+2)$-brane charge and $p$-brane charge the same
as that of a $p$-brane on $Z$.  This suggests that the system could be deformed
to a system consisting just of a $p$-brane wrapped on $Z$.  

As evidence for this, we note that the tachyon field of the brane-antibrane
pair, because it has charges $(1,-1)$ under the $U(1)\times U(1)$ that live
on the brane and antibrane, should be a section of ${\cal L}$.
Hence we can take
\eqn\jsn{T=c\cdot s,}
with $c$ a constant chosen so that far from $Z$, $|T|$ is equal to its
vacuum expectation value.  
In future, we will generally omit constants analogous to $c$, to avoid
cluttering the formulas.
The basic assumptions about brane-antibrane
annihilation then suggest that with this choice of $ T$, the system is
in a vacuum state except near $Z$ and can be described by a $p$-brane
wrapped on $Z$.

\bigskip\noindent{\it Incorporation Of Lower Charges}

A fuller description actually requires the following generalization.
Note that a $p$-brane wrapped on $Z$ has in general in addition to its
$p$-brane charge also $r$-brane charges
with $r=p-2,p-4,\dots$.  Moreover, these depend on the choice of a line bundle
${\cal M}$ on $Z$.  Thus, to fully describe all states with a $p$-brane
wrapped on $Z$ in terms of states of a brane-antibrane pair wrapped
on $Y$, we need a way to incorporate ${\cal M}$ in the discussion.

If ${\cal M}$ extends over $Y$, we incorporate it in the above discussion
just by placing the line bundle ${\cal L}\otimes {\cal M}$ on the
$p+2$-brane and the line bundle ${\cal M}$ on the $\bar {p+2}$-brane.
The tachyon field  $T$, given its charges $(1,-1)$, is a section
of $(\L\otimes \M)\otimes \M^{-1}={\cal L}$, so we can  take
$T= s$ and flow (presumably) to a configuration containing
only a $p$-brane wrapped on $Z$.  The $r$-brane charges with $r<p$ now
depend on $\M$ in a way that has a simple interpretation:
on the $p$-brane worldvolume there is a $U(1)$ gauge field with
line bundle $\M$.

More generally, however, $\M$ may not extend over $Y$.  To deal with
this case, we need to use another of the basic constructions of K-theory.
First we describe it in mathematical terms.
Let $Z$ be a submanifold of a manifold $Y$, and $Z'$  a tubular neighborhood
of $Z$ in $Y$ (this means that we pick a suitable metric on $Y$, and let
$Z'$ consist of points of distance $<\epsilon$ from $Z$, for some small
$\epsilon$). 
Let $\bar Z$ be the closure of $Z'$ (the points of distance $\leq \epsilon$
from $Z$)
and $Z^*$ its boundary (the points of distance precisely $\epsilon$).  
Suppose that $E$ and $F$ are two bundles over
$Z$ of the same rank (in our example so far, they are line bundles,
$E=\L\otimes \M$ and $F=\M$), so that the pair $(E,F)$ defines an element
of $\K(Z)$ .  Pull $E$ and $F$ back to $\bar Z$, so that $(E,F)$ defines
an element of $\K(\bar Z)$.
The tachyon field $T$, which is a section of $ E\otimes F^*$,
 can be regarded as a bundle map
\eqn\kido{T:F\to E.}
Suppose that $T$ is  a tachyon field on $\bar Z$ which (when
viewed in this way as a bundle map)
 is an isomorphism (an invertible map)
 if restricted to $Z^*$.   Then from this data,
one can construct an element of $\K(Y)$.\foot{Moreover, the construction
gives a natural map from elements of $\K(\bar Z)$ trivialized on $ Z^*$
to $\K(Y)$, in the sense that the image of $(E,F)$ with bundle map
$T$ is, for any $H$,
 the same as the image of $(E\oplus H,F\oplus H)$, with bundle
map $T\oplus 1$.}
The construction is made as follows.  
Let $Y'=Y-Z'$; thus $Y'$ consists of $Z^*$ and its ``exterior'' in $Y$.
If we could extend the bundle $F$ from $Z^*$ over all of $Y'$
then $F$ would be defined over all of $Y$ (since it is defined already
on $\bar Z$).  Since $E$ is isomorphic to $F$ on $Z^*$ (via $T$), we could
extend it over $Y'$ by declaring that it is isomorphic to $F$ over $Y'$.
The pair $(E,F)$ of bundles on $Y$ then give the desired element of
$\K(Y)$.  

If $F$ does not extend over $Y'$, one proceeds as follows.
By a standard lemma in K-theory (Corollary 1.4.14 in \atiyah), 
there is a bundle $H$ over $Z$ such that
$F\oplus H$ is trivial over $Z$, 
and hence is trivial when pulled back to $\bar Z$.
Replacing $E, F$, and $T$ by $E\oplus H$, $F\oplus H$, and $T\oplus 1$,
we can extend $F\oplus H$ over $Y$ (as a trivial bundle), and extend
$E\oplus H$ by setting it equal to $F\oplus H$ over $Y'$. The
pair $(E\oplus H, F\oplus H)$ then give the desired element of $\K(X)$.  Note
that $E\oplus H$ and $F\oplus H$ are isomorphic over $Y'$ but not over $Y$.

This construction is precisely what we need to express in terms of 
$(p+2)$-branes
on $Y$ a $p$-brane on $Z$ that supports a line bundle $\M$.
We find a bundle $H$ over $Z$ such that $\M\oplus H$ is trivial (and so
extends over $Y$).  
$\L\otimes\M\oplus H$ is extended over $Y$ using the fact that (via $T\oplus 
1$)
it is isomorphic to $\M\oplus H$ away from $Z$.
Then we consider a collection of $(p+2)$-branes
on $Y$ with gauge bundle $\L\otimes \M\oplus H$, and $(\bar{p+2})$-branes
on $Y$ with gauge bundle $\M\oplus H$.  
The number of branes of each kind is 1 plus the rank of $H$.
The tachyon field is $T\oplus 1$
 near $Z$, and is in the gauge orbit of the vacuum outside of $Z'$.
The system thus describes, under the usual assumptions, a $p$-brane on $Z$
with gauge bundle $\M$.

In a similar fashion, we could have started with any collection of $n$
$p$-branes wrapped on $Z$, with $U(n)$ gauge bundle ${\cal W}$, and expressed
it in terms of a collection of $(p+2)$-branes and antibranes on $Y$.
One pulls back ${\cal W}$ to $\bar Z$, uses the tachyon field
$\tilde T=T\otimes 1$ to identity $ {\cal W}$ with $\L\otimes {\cal W}$
on the boundary of $\bar Z$, and then uses the pair of bundles
 $(\L\otimes {\cal W},
{\cal W})$ 
 to determine a class in $\K(Y)$.  Such a class is, finally,
interpreted in terms of a collection of branes and antibranes wrapped on $Y$.
The $p$-brane charge is $n$; the $r$-brane charges for $r<p$ depend on 
${\cal W}$.

\subsec{Spinors And The Anomaly}

Specializing to the case that $Y$ coincides with the spacetime manifold $X$
and $Z$ is of codimension two in $X$, this construction shows that
whatever can be done with sevenbranes can be done with ninebranes.
We now wish to show that brane wrapping on a submanifold $Z$ of codimension
greater than two can likewise be expressed globally
in terms of ninebranes.

Under favorable conditions, there might be a chain of embeddings
$Z\subset Z'\subset \dots \subset X$, with codimension two at each stage.
Then we could inductively use the construction already explained.  
In general, however, such a chain of embeddings 
will not exist globally.
Instead, we will use spinors, building on facts explained at the end
of section 4.1.

\def\SS{{\cal S}}
Let $N$ be the normal bundle to $Z$ in $X$.  If
$Z$ has codimension $2k$, then the   structure group of $N$ is $SO(2k)$.
We suppose first that $N$ is a spin bundle, which means that $w_2(N)=0$
and that there are bundles $\SS_+$, $\SS_-$ associated to $N$ by using
the positive and negative chirality spin representations of $SO(2k)$.
\foot{$N$ may have different spin structures.  In fact, since to do Type IIB
theory on $X$, $X$ is endowed with a spin structure, a choice of spin
structure on $N$ is equivalent to a choice of spin structure on $Z$.
The K-theory class determined by $Z$ may in general depend on its spin
structure -- or  more generally, on its ${\rm Spin}_c$ structure,
as described below.}
We consider a system of $2^{k-1}$ 9-branes, and the same number of
$\bar 9$ branes, with the gauge bundles on them, near $Z$, being 
$\SS_+$ and $\SS_-$.\foot{We tacitly assume for the moment that $\SS_\pm$ 
extend
over $X$ and postpone the technicalities that arise when this is not so.} 
This system has all $r$-brane
charges zero for $r>p$ (as  they cancel between the branes and antibranes),
while its $p$-brane charge is that of a single brane wrapped on $Z$.

The tachyon field $T$ from the $9 $-$ \bar 9$ sector
should be a map from $\SS_-$ to $\SS_+$.  The basic such maps are the Dirac
Gamma matrices $\Gamma$.
  We identify a tubular neighborhood $Z'$ of $Z$ in $X$ with the 
vectors in $N$ of length $<1$, and for $x\in Z'$, we write
\eqn\iko{T=\vec \Gamma\cdot \vec x.}
$T$ gives a unitary isomorphism between $\SS_-$ and $\SS_+$ on the boundary of
$Z'$ (since $\vec\Gamma\cdot \vec x$ is unitary if $\vec x$ is a unit vector),
so (after scaling by a constant $c$, which we suppress) $T$ on the boundary
of $Z'$ is in the gauge orbit of the vacuum.
If, therefore, we can extend this configuration over $X$, keeping $T$
equal to its vacuum expectation value, then we will get a system
of 9-branes and $\bar 9$-branes that represents a single $p$-brane
wrapped on $Z$.  As in the discussion at the end of section 4.2,
the configuration we have described can be extended over $X$ if the bundle
$\SS_-$ so extends.  Otherwise, we pick a suitable $H$ such that $\SS_-\oplus
H$ extends, and replace $(\SS_+,\SS_-)$ by $(\SS_+\oplus H,\SS_-\oplus H)$
and $T$ by $T\oplus 1$. 

More generally, to describe a $p$-brane on $Z$ with a line bundle
$\M$, we use the 9-brane configuration 
\eqn\kindo{(\M\otimes \SS_+,\M\otimes \SS_-), }
or still more generally
$(\M\otimes \SS_+\oplus H,\M\otimes \SS_-\oplus H)$, with $H$ chosen
so that these bundles extend over $X$.  The tachyon field is still $T=
\vec \Gamma\cdot \vec x$ (or $\vec\Gamma\cdot\vec x\oplus 1$)
in a neighborhood of $Z$, and lies in its
vacuum orbit in the 
complement of this neighborhood.

\bigskip\noindent{\it The ${ Spin}_c$ Case}

So far we have assumed that the normal bundle $N$ to $Z$ in spacetime
is spin, $w_2(N)=0$.  What if it is not?

If instead of being spin, $N$ admits a ${\rm Spin}_c$ structure,
then we can proceed much as before.  
$N$ not being spin means the following.  If we cover $X$ with open sets
$W_i$, then the would-be transition functions $w_{ij}$ of $\SS_+$
(or similarly $\SS_-$) on $W_i\cap W_j$ obey 
\eqn\gifo{w_{ij}w_{jk}w_{ki}=
\phi_{ijk},} where $\phi_{ijk}=\pm 1$.  The $\phi_{ijk}$ are a two-cocycle
(with values in $\{\pm 1\}$) defining $w_2(N)\in H^2(Z,\Z_2)$.
\foot{The cocycle property can be proved from \gifo\ as follows.
First rewrite this formula as $w_{ij}w_{jk}=\phi_{ijk}w_{ik}$.
Now consider the product $w_{ij}w_{jk}w_{kl}$.  This product
can be evaluated by associativity as $(w_{ij}w_{jk})w_{kl}
=\phi_{ijk}\phi_{ikl}w_{il}$, or $w_{ij}(w_{jk}w_{kl})
=\phi_{ijl}\phi_{jkl}w_{il}$.  Comparing these gives the cocycle
relation $\phi_{ijk}\phi_{ikl}=\phi_{ijl}\phi_{jkl}$.}
$N$ being ${\rm Spin}_c$ means that there is a line bundle ${\cal L}$ over
$Z$, with the following property.  Let $f_{ij}$ be transition functions
on $W_i\cap W_j$ defining $\L$.  The ${\rm Spin}_c$ property arises
if a square root $\L^{1/2}$ of $\L$ does not exist as a line bundle,
but is obstructed by the same cocycle that obstructs existence of $\SS_\pm$.
This happens
if putative transition functions $g_{ij}=\pm \sqrt{f_{ij}}$ of $\L^{1/2}$
(with suitable choices of the signs) obey $g_{ij}g_{jk}g_{ki}=\phi_{ijk}$.
In this case, the cocycle cancels out in the transition functions
$g_{ij}w_{ij}$ of the vector
bundles $\L^{1/2}\otimes \SS_\pm $, and these objects 
(which are sometimes called ${\rm Spin}_c$ bundles) exist as honest vector
bundles, even though the factors $\L^{1/2}$ and $\SS_\pm$ do not separately
have that status.

Notice that such an $\L$, if it exists, will generally be far from unique.
We could pick any line bundle $\M$ over $Z$ and replace $\L$ by
$\M^2\otimes \L$; this maps the ${\rm Spin}_c$ bundles $\L^{1/2}\otimes
\SS_\pm$ to $\M\otimes \L^{1/2}\otimes \SS_\pm$, which certainly exist
if and only if $\L^{1/2}\otimes \SS_\pm $ do.  One way to characterize
the allowed $\L$'s is as follows.  Define $x\in H^2(Z,\Z)$ by
$x=c_1(\cal L)$.  Then modulo 2, $x$ is invariant under $\L\to \M^2\otimes \L$,
and it can be shown that existence, in the above sense, of 
$\L^{1/2}\otimes \SS_\pm$ is equivalent to
\eqn\eqtopol{x\cong w_2(N)\,\,{\rm mod}\,2.}
The criterion that a ${\rm Spin}_c$ structure
  exists can be stated as follows.
Consider the exact sequence
\eqn\kilop{0\to \Z\underarrow{2}\Z\to \Z_2\to 0,}
where the first map is multiplication by 2 and the second is reduction
modulo 2.  The associated long exact sequence of cohomology groups
reads in part
\eqn\retup{\dots \to H^2(Z,\Z)\to H^2(Z,\Z_2)\underarrow{\beta} 
H^3(Z,\Z)\to\dots.}
The image of $w_2(N)\in H^2(Z,\Z_2)$ under the map that has been called $\beta$
in \retup\ is an element of $H^3(Z,\Z)$
called $W_3(N)$.  ($\beta$ is called the Bockstein homomorphism.)
Exactness of the sequence \retup\ implies that $w_2(N)$ can be lifted
to $x\in H^2(Z,\Z)$ -- and hence $N$ is ${\rm Spin}_c$ -- if and only
if $W_3(N)=0$.

Returning now to our overall problem  of interpreting a brane wrapped
on $Z$ in terms of an element of $\K(X)$, if the bundle $N$ is ${\rm Spin}_c$
we can proceed precisely as we did in the spin case.  
The bundles $(\L^{1/2}\otimes \SS_+,\L^{1/2}\otimes \SS_-)$, with the
tachyon field $T$ still defined as in \iko, determine the desired
element of $\K(X)$ that represents a brane wrapped on $Z$.  The possibility
of tensoring $\L^{1/2}$ with an arbitrary line bundle $\M$ just corresponds
to the fact that the brane wrapped on $Z$ could support an arbitrary line
bundle.

\bigskip\noindent{\it The Topological Obstruction}

Now we come to a key point.  What if the normal bundle $N$ is {\it not}
${\rm Spin}_c$?

There seems to be a puzzle.  If $N$ is not ${\rm Spin}_c$, 
a brane wrapped on $Z$ does not determine a K-theory class.  This appears
to contradict the relation between branes and K-theory.

The answer, surprisingly, is that if $N$ is not ${\rm Spin}_c$,
a brane cannot be wrapped on
$Z$.  This follows from a topological obstruction to brane wrapping
that was observed in a particular situation in \witten\ and
has been extracted from world-sheet global anomalies \freedwitten.
The obstruction in question appears in eqn.  (3.13) of \witten.
(The $[H]$ term in that equation can be dropped, since we are assuming
at the present that the    cohomology class of the Neveu-Schwarz
three-form is zero.)  Let $w_i$, $i=1,2,\dots $ denote Stieffel-Whitney
classes.  In particular, let $w_i(Z)$ be the Stieffel-Whitney classes
of the tangent bundle of $Z$.
 Let $W_3(Z)=\beta(w_2(Z))$ (with $\beta$ being the Bockstein).  
 Equation (3.13) of \witten\ says that
a brane can wrap on $Z$ if and only if $W_3(Z)=0$.

Using the fact that
 $X$ is spin, $w_1(X)=w_2(X)=0$, and that $Z$ is orientable, $w_1(Z)=0$,
a standard argument\foot{The multiplicativity of the total Stieffel-Whitney
class in direct sums gives $(1+w_1(X)+w_2(X)+\dots)=(1+w_1(Z)+w_2(Z)+\dots)
(1+w_1(N)+w_2(N)+\dots)$.  With $0=w_1(X)=w_2(X)=w_1(Z)$, we get
$0=w_1(N)$ and $w_2(N)=w_2(Z)$.} gives $w_2(N)=w_2(Z)$, and hence
$W_3(N)=W_3(Z)$.   This is compatible with the idea that the charges
carried by branes are measured by K-theory classes.  If $W_3(N)\not= 0$,
then the construction of a K-theory class that would be carried by a brane
wrapped on $Z$ fails, as we have seen above, but this presents no problem
since such a wrapped brane does not exist.

Even when $W_3(N)$ is nonzero,  it is  possible to have
a configuration consisting of {\it several} branes wrapped on $Z$, supporting
suitable gauge fields.  The gauge bundle ${\cal W}$ on $Z$ must
not be a true vector bundle; its transition functions must close up to $\pm$
signs in just such
a way that ${\cal W}\otimes \SS_\pm$ exists as a vector bundle.
The most obvious choice is ${\cal W}=\SS_+$ (or $\SS_-$); the bundles
$\SS_+\otimes \SS_\pm$ exist as they can be expressed in terms of
differential forms.  If $Z$ is of codimension $2k$ in $X$, then
the rank of $\SS_+$ is $2^{k-1}$.  The K-theory class associated with
$(\SS_+\otimes \SS_+,\SS_+\otimes \SS_-)$ and the usual tachyon field
\iko\ describes a configuration of $2^{k-1}$ branes wrapped on $Z$, 
supporting a  ``gauge bundle'' $\SS_+$.    In some cases with
$W_3(N)\not=0$, it is possible to find a more economical solution
with a smaller (but  even) number of branes wrapped on $Z$.
 
I will not in this paper discuss the analogous issues for Type IIA,
except to note that having come to this point,
the reader may now find the
comments in the last paragraph of
section three to be clearer.  We move on next to discuss some analogous
questions for Type I superstrings.

\subsec{Spinors And Type I Branes}

In studying Type I superstrings, we begin with the $D$-string.
We wish to exhibit it as a bound state of 9-branes and $\bar 9$-branes.
 
We want to describe a $D$-string located at $x^1=\dots =x^8=0$ in $\R^{10}$;
its world-volume $Z$ is parametrized by $x^0$ and $x^9$. 

The group of rotations keeping $Z$ fixed is $K=SO(8)$.  $K$ rotates
the eight-vector $\vec x=(x^1,\dots, x^8)$.  The two
spinor representations of $K$ are both eight-dimensional; we call
them $S_+$ and $S_-$.  We regard the $\Gamma$-matrices $\Gamma^i$ as 
maps from $S_-$ to $S_+$.

We consider a configuration of eight 9-branes and eight $\bar 9$-branes.
We take the gauge bundles on these branes to be trivial,
but we take the rotation group $K$ to act on the Chan-Paton labels,
with the rank eight bundle of the 9-branes transforming as $S_+$, and
the $\bar 9$-brane bundle transforming as $S_-$.

For the tachyon field, we take
\eqn\idno{T(\vec x)=f(|\vec x|)\vec \Gamma\cdot \vec x,}
with $f$ a function that is 1 near $|\vec x|=0$, and 
$c/|\vec x|$ for $|\vec x|\to\infty$,
with a suitable constant $c$.  $T/c$ is an orthogonal matrix for $|\vec x|\to
\infty$; $c$ is chosen so that $T$
 lies in the gauge orbit of the vacuum.  In the spirit
of Sen's constructions, we expect that this configuration is equivalent
to the vacuum except near $\vec x=0$.  Thus it describes a string localized
near $\vec x=0$.  In future we will generally omit the $f(|\vec x|)$ factor
to avoid clutter.

Note that with the chosen action of $SO(8)$ on the
gauge bundles, the tachyon field is $SO(8)$-invariant, so the construction
is manifestly $SO(8)$-invariant.  Moreover, if we restore the extra 32
9-branes that are needed for Type I tadpole cancellation, then $SO(32)$
gauge symmetry of these 9-branes is just a spectator in this construction;
the
construction is manifestly $SO(32)$-invariant.  

In this construction,
the gauge field on the branes must be chosen so that $T$ is covariantly
constant near infinity.  
Since $\vec x\to \vec\Gamma\cdot \vec x/|x|$ is
the generator of $\pi_7(SO(8))=\Z$, the configuration that we have built
is a ``gauge string,'' in the sense of section 2.  However, now we have
used extra brane-antibrane pairs to enlarge the gauge group, and have made
the construction in a manifestly $SO(32)$-invariant way.

Now, as in section 4.3, we would like to make this construction globally.
This involves no essential novelty compared to Type IIB, except perhaps
for the fact that there is no topological anomaly to worry about.
We consider a two-surface $Z$ in a Type I spacetime $X$. In Type I
superstring theory, $X$ is spin, so $w_1(X)=w_2(X)=0$.  To wrap a 
$D$-string on $Z$, $Z$ must be orientable, so $w_1(Z)=0$, and hence
(as orientable two-manifolds are spin) $w_2(Z)=0$.  
We pick a spin structure on $X$ (since Type I requires a spin structure)
and on $Z$ ($D$-string wrapping on a two-cycle is expected to depend on
a choice of spin structure on the two-cycle).  Let $N$ be the normal
bundle to $Z$ in $X$.
Since $(1+w_1(X)+w_2(X)+\dots)=(1+w_1(Z)+w_2(Z)+\dots)(1+w_1(N)+w_2(N)+\dots)$,
one has $w_1(N)=w_2(N)=0$, so spin bundles
$\SS_-$ and $\SS_+$ (derived from $N$ using the spin representations of 
$SO(8)$)
exist.  More specifically, the chosen spin structures for $X$ and $Z$ determine
in a natural way a spin structure for $N$.

  Taking eight 9-branes whose gauge bundle near $Z$ is identified
with $\SS_+$, and eight $\bar 9$-branes with gauge bundle near $Z$
identified with $\SS_-$, and a tachyon field that looks like 
$T=\vec \Gamma\cdot \vec x$ near $Z$, we express the $D$-string wrapped on
$Z$ in terms of eight $9$-$\bar 9$ pairs.  An interesting point is that
as $Z$ is two-dimensional and
$w_1(\SS_-)=w_2(\SS_-)=0$, $\SS_-$ is actually trivial along $Z$, and hence
can be extended over $X$ as a trivial bundle.  Hence, in contrast to Type IIB,
there is no need to ``stabilize'' by adding extra $9$-$\bar 9$ pairs;
eight of them is always enough.

There was no need in this construction to assume that $Z$ is connected.
So any collection of $D$-strings, at least if they are disjoint, 
can be represented by a configuration
of eight $9$-$\bar 9$ pairs.  There is no need to introduce eight
more pairs for every $D$-string!  Since $D$-strings are equivalent to
perturbative heterotic strings, this is close to saying that the
second quantized Fock space of perturbative heterotic strings can be
described by configurations of eight $9$-$\bar 9$ pairs.

\bigskip\noindent{\it Fivebranes}

We will briefly discuss fivebranes in a similar spirit. 

The basic local fact making it possible to interpret
 fivebranes as ninebrane configurations
is that a Type I fivebrane is equivalent to an instanton on the
ninebranes that fill the vacuum \ref\uwitten{E. Witten, ``Small Instantons
In String Theory,'' Nucl. Phys. {\bf B460} (1996) 541,
hep-th/9511030.}.  If one nucleates extra $9$-$\bar 9$ pairs,  there
is more flexibility.   Consider a system of four $9$-$\bar 9$
pairs, so that the 9-brane Chan-Paton group is $SO(4)=SU(2)\times SU(2)$.
Place an instanton of instanton number 1 in one of the $SU(2)$'s.
This makes a configuration whose fivebrane number is equal to 1.
(We picked four $9$-$\bar 9$ pairs as it is the smallest number for
which the fivebrane number can be 1.)  With a suitable tachyon field
(very similar in fact to what is discussed in \senthree\ in showing
that Type I $D$-strings can be made from fivebranes), this should
be equivalent to a fivebrane.

I leave it to the reader to analyze this construction globally, and show
that there is no obstruction to similarly making a fivebrane that is wrapped
on any spin six-cycle out of $9$-$\bar 9$ pairs. 

The $SO(4)\times SO(4)$ gauge symmetry of four $9$-$\bar 9$ pairs
is broken to a diagonal $SO(4)$ by the tachyon field and to $SU(2)={\rm Sp}(1)$
by the instanton.  This is the usual ${\rm Sp}(1)$ gauge symmetry of
a Type I fivebrane.  The ${\rm Sp}(1)$ gauge symmetry is associated
with the following mathematical fact.  Type I fivebrane charge
takes values in $\KO(\S^4)$ (or equivalently $\KO(\R^4)$ with compact support,
the $\R^4$ parametrizing here the directions normal to the fivebrane).
Type IIB fivebrane charge likewise takes values in $\K(\S^4)$.  The
groups $\KO(\S^4)$ and $\K(\S^4)$ are both isomorphic to $\Z$.  But
the natural map from $\KO(\S^4)$ to $\K(\S^4)$ (defined by forgetting that
the bundles are real), is multiplication by 2.  This is because the generator
of $\K(\S^4)$ is an $SU(2)$ instanton field, which is a pseudoreal bundle
and  must be embedded in
$SO(4)$ (as we did two paragraphs ago)  if one
wants to make it real.  The embedding in $SO(4)$ doubles the charge,
so the fivebrane charge of
a Type I fivebrane is twice that of a Type IIB fivebrane.

This completes the demonstration that the Type I configurations
built from the usual supersymmetric branes ($p$-branes for $p=1,5,9$)
 represent classes in $\KO(X)$.  However, there is more to say.
The relation to $\KO(X)$ suggests that Type I should also have, for example,
zerobranes -- associated with $\KO(\S^9)=\Z_2$ -- and $-1$-branes
-- associated with $\KO(\S^{10})=\Z_2$.  
Concretely, the assertions that ${\rm KO}(\S^9)=\Z_2$
and that ${\rm KO}(\S^{10})=\Z_2$ are equivalent to the assertions
that $\pi_8(SO(k))=\pi_{9}(SO(k))=\Z_2$ for sufficiently large $k$.
We examined topological defects associated with these homotopy groups
in section 2; we will now reexamine them in light of our experience
with K-theory.

\subsec{The Type I Zerobrane and $-1$-Brane}

For $n=9$ or 10, we will think of ${\rm KO}(\S^n)$ as $\KO(\R^n)$
with compact support.  An element of $\KO(\R^n)$ with
compact support is described by giving two $SO(k)$ bundles $E$ and $F$ over
$\R^n$ (for some $k$), with a bundle map $T:F\to E$ that is an isomorphism
near infinity.  The physical interpretation, as we have seen, is that
$E$ is the Chan-Paton bundle of $k$ 9-branes, $F$ the Chan-Paton bundle
of $k$ $\bar 9$-branes, and $T$ the $9$-$\bar 9$ tachyon field.  Since
$\R^n$ is contractible, the bundles $E$ and $F$ are trivial; the topology
is all in the ``winding'' of $T$ near infinity.

The standard mathematical descriptions \abs\ 
of generators of ${\rm KO}(\R^9)$
and ${\rm KO}(\R^{10})$ with compact support
are similar to what we have seen already
in describing the more familiar supersymmetric branes of Type II and
Type I superstring theory.
In each case, we   take $E$ and $F$ to be trivial bundles on which the
rotation group of $\R^9$ or $\R^{10}$ acts in the spin representation.
Since we want KO theory, we must use real spin representations.  

In the case of $\R^9$, we need the spinor representation of $SO(9)$.
There is only one such irreducible representation $S$. 
It is real and of dimension 16, so we consider the case that $E$ and 
$F$ are 16-dimensional and transform under rotations like $S$.
The tachyon field is given by the familiar formula:
\eqn\pollyp{T(x)=\sum_{\mu=1}^9\Gamma_\mu x^\mu,}
with $x^\mu,\,\mu=1,\dots, 9$ 
the coordinates of $\R^9$, and $\Gamma_\mu$ the $\Gamma$ matrices.
\foot{As earlier, the formula for $T(x)$ should more properly be
$T=f(|x|)\Gamma_\mu x^\mu$, with $f$ constant for small $|x|$ and
$f\sim 1/|x|$ for large $|x|$.  To keep the formulas simple, we omit
this factor.}
This configuration is manifestly $SO(9)$-invariant, with the indicated
action of $SO(9)$ on the Chan-Paton bundles $E$ and $F$.

Now to compare this to the description of the Type I zerobrane
given in \senthree, we want to make an $8+1$-dimensional split
of the   coordinates and Gamma matrices.  We pick an $SO(8)$ subgroup of
$SO(9)$, under which the $x^\mu$ break up as $\vec x, x^9$,
with $\vec x=(x^1,\dots, x^8)$ and $x^9$ the last coordinate.
The representation $S$ of $SO(9)$ breaks up under $SO(8)$ as 
$S_+\oplus S_-$, with $S_+$ and $S_-$ the positive and negative
chirality spinor representations of $SO(8)$, which are of course
both real.  We write the $SO(8)$ Gamma matrices as 
$\vec \Gamma:S_-\to S_+$, $i=1,\dots ,8$,
and  their transposes $\vec \Gamma^T:S_+\to S_-$.
In a basis in which we write the $SO(9)$ spinors as
\eqn\writet{S=\left(\matrix{ S_+\cr S_-\cr}\right),}
the tachyon field of equation \pollyp\ is then
\eqn\nudnik{T=\left(\matrix{ x^9 & \vec \Gamma\cdot \vec x\cr
                             \vec\Gamma^T\cdot \vec x & -x^9\cr}\right).}
If we make a change of basis on the $\bar 9$-branes by the
matrix
\eqn\plp{M=\left(\matrix{ 0 & 1 \cr 1 & 0 \cr}\right),}
then the tachyon field is transformed to 
\eqn\udnik{T=\left(\matrix{\vec \Gamma\cdot \vec x & x^9\cr
                              -x^9               & \vec\Gamma^T\cdot \vec x\cr}
                              \right).}
This formula has a nice intuitive interpretation.  Suppose first
that we neglect the off-diagonal blocks in \udnik.  Then
the system splits up into two decoupled systems each containing
eight $9$-$\bar 9$ pairs.  The first set of eight pairs has tachyon field
\eqn\conc{T_1=\vec\Gamma\cdot \vec x}
and the second has tachyon field
\eqn\ponc{T_2=\vec \Gamma^T\cdot \vec x.}
As we have discussed at the beginning of the present section, $T_1$
describes a $D$-string located at $x^1=\dots=x^8=0$.  Since $T_2$ is made
from $T_1$ by exchanging the 9-branes with $\bar 9$-branes, $T_2$
describes  an anti $D$-string located at $x^1=\dots = x^8=0$.
This is precisely the configuration of a coincident $D$-string and
anti $D$-string coinsidered by Sen \senthree.  Moreover, the off-diagonal
blocks in \udnik\ can be understood as a tachyon field which connects the
$D$-string and anti $D$-string and is odd under $x^9\to -x^9$.  This
is precisely the solitonic configuration of the $1$-$\bar 1$ tachyon field
that is described in \senthree.  So we have made contact with this
form of Sen's construction.

In section 6, we will attempt to give a slightly simplified version of
the worldsheet description in \senfour.  As preparation for that, let
us notice the following suggestive fact.  To describe Type IIB branes
of codimension $2k$, we used spinors of $SO(2k)$ of definite chirality.
The dimension of the chiral spinors of $SO(2k)$ is $2^{k-1}$,
and this was the number of $9$-$\bar 9$ pairs used to describe a
brane of codimension $2k$.

For the Type I zerobrane, the codimension is 9.  To use the same formula
as in the other cases, we would set $2k=9$ and expect the number
of $9$-$\bar 9$ pairs to be $2^{k-1}=8\sqrt 2$.  This does not make
sense as it is not an integer. The actual number
in the above construction is 16, larger than $8\sqrt 2$ by a factor
of $\sqrt 2$.  We will seek to interpret the extra
factor of $\sqrt 2$ in section 6.

\bigskip\noindent{\it The $-1$-Brane}

Now we consider the $-1$-brane.  For this, we must take $E$ and $F$ to
be spinor representations of $SO(10)$.  Moreover, we must use {\it real}
spinor representations, as we are doing Type I superstrings and KO theory.
The group $SO(32)$ has a unique irreducible real spinor representation
$S$; it is 32-dimensional.  The $-1$-brane is described with 32 $9$-$\bar 9$
pairs and a tachyon field given by the usual formula $T=\vec \Gamma\cdot\vec 
x$.

As preparation for a worldsheet construction discussed in section six,
we note the following.  Although the representation $S$ is irreducible
over the real numbers, if complexified it decomposes as $S=S_+\oplus S_-$
where $S_+$ and $S_-$ are the 16-dimensional complex spinor
representations of $SO(10)$ of positive and negative chirality.
The tachyon field $T=\vec\Gamma\cdot \vec x$
 of course reverses the chirality.  If therefore
we were working in Type IIB and all matrices were complex, we would
decompose this system as a sum of two subsystems, one with $9$-brane
and $\bar 9$-brane representations
$(E_1,F_1)=(S_+,S_-)$, and the second with representations $(E_2,F_2)
=(S_-,S_+)$.  (The tachyon field is $T=\vec\Gamma\cdot \vec x$, mapping
$F_1$ to $E_1$ and  $F_2$  to $E_2$.)
The $(S_+,S_-)$ system is the by now familiar description of 
a Type IIB $-1$-brane, and the $(S_-,S_+)$ system,
 which has the chiralities or equivalently
the 9-branes and $\bar 9$-branes reversed, describes similarly
 a Type IIB anti $-1$-brane.
The orientation projection that reduces Type IIB to Type I acts by
complex conjugation, so it exchanges $S_+$ and $S_-$ and hence exchanges
the $-1$-brane with the anti $-1$-brane.  This information will enable
us in section six to give a worldsheet description of the Type I
$-1$-brane.

One could also make a $9+1$-dimensional
or $8+2$-dimensional split of the 10-dimensional
Gamma matrices, and describe the Type I $-1$-brane in terms of 
configurations of zerobranes or onebranes and coincident antibranes,
with suitable tachyon fields.  We will omit this.

\bigskip\noindent{\it The Sevenbrane and Eightbrane}

Both in section two and here, we have focussed our discussion of Type I
on the topological objects associated with $\pi_7$, $\pi_8$, and $\pi_9$.
What other nonzero homotopy groups are there in a range that is relevant
in ten dimensions?  $\pi_3$ is nonvanishing and is associated with the
familiar Yang-Mills instantons, or alternatively with the Type I fivebrane,
which we have discussed in section 4.4.  The other candidates are
\eqn\hsx{\pi_0(O(32))=\pi_1(O(32))=\Z_2,}  
where here we recall that the perturbative gauge group is more nearly
$O(32)$ than $SO(32)$.\foot{Since 
every open string has two ends, the generator $-1$ of the center
of $O(32)$ acts trivially on all open string states.  The perturbative
gauge group
as opposed to the group acting on the Chan-Paton factors at the end
of a string is thus $O(32)/\Z_2$. If one replaces
$O(32)$ by $O(32)/\Z_2$, one gets an extra $\Z_2$ in $\pi_1$.  This
leads to the possibility of considering bundles without
``vector structure,'' a generalization we will make in section five.}
 This suggests that
one could make in Type I a sevenbrane and an eightbrane,
related to $\KO(\S^2)$ and $\KO(\S^1)$ respectively.  

As usual we identify $\KO(\S^n)$ with $\KO(\R^n)$ with compact support;
and we describe an element of $\KO(\R^n)$ with compact support by
giving trivial bundles $E,F$ on $\R^n$ and a tachyon map between them
that is invertible at infinity.  The formula for the tachyon map
is always $T=\vec \Gamma\cdot \vec x$.

For $n=1$, there is only        one Gamma matrix.  We can take it
to be the $1\times 1$ unit matrix.  The tachyon field is thus
\eqn\oxno{T=x^9}
(times a convergence factor such as $1/\sqrt {1+(x^9)^2}$) 
where for convenience we have labeled as $x^9$
 the coordinate on $\R^1$.
Thus, the eightbrane is a ``domain wall,'' located at $x^9=0$.
It is constructed from a single $9$-$\bar 9$ pair, with a tachyon field
that is positive on one side and negative on the other.  I would conjecture
-- but will not try to prove here --
that the sign of the $-1$-brane amplitude is reversed in crossing this
domain wall.

The sevenbrane is similarly constructed with two $9$-$\bar 9$ pairs
and $2\times 2$ real $\Gamma$ matrices.  I would conjecture, but will
again not try to prove, that a zerobrane wavefunction picks up a factor
of $-1$ under parallel transport about the sevenbrane.

These conjectures assert that there is a sort of discrete electric-magnetic
duality between $-1$-branes and 8-branes, and between  0-branes and 7-branes.
Recall that in ten dimensions, dual $p$-branes and $q$-branes  carrying
additive charges obey $p+q=6$.  In the case of branes carrying
discrete charges, one apparently has $p+q=7$.  
                             
\bigskip\noindent{\it A Note on Bott Periodicity}

Finally, we make a note on Bott periodicity, which asserts that
for KO-theory with compact support, one has $\KO(\R^n)=\KO(\R^{n+8})$.
In particular, Bott periodicity maps the $-1$-brane to the 7-brane
and the 0-brane to the 8-brane.

The periodicity map can be described as follows.  Consider an element
of $\KO(\R^n)$ described by trivial bundles $(E_0,F_0)$ on $\R^n$
with a tachyon map $T_0:F_0\to E_0$.  From this data one constructs
an element of $\KO(\R^{n+8})$ by letting $S_+$ and  $S_-$ be the chiral
spinor representations of $SO(8)$, and setting $E=E_0\otimes (S_+\oplus S_-)$,
$F=F_0\otimes (S_+\oplus S_-)$.  We also denote by $\vec \Gamma:S_-\to S_+$
the $SO(8)$ Gamma matrices, and by $\vec x$ the last eight coordinates of
$\R^{n+8}$. Then
in a hopefully evident notation one takes the tachyon field to be
\eqn\jcn{T=\left(\matrix{T_0 & \vec\Gamma \cdot\vec x \cr 
         \vec \Gamma^T\cdot \vec x & - T_0}\right).}
Comparing to \oxno\ and \nudnik, we see that the relation between the
eight-brane and the zerobrane is a typical example of this periodicity
map.

\newsec{Some Generalizations}

In this section, we consider three types of generalization of
the above discussion, involving orbifolds, orientifolds, and
the incorporation of the Neveu-Schwarz three-form field $H$.

\subsec{Orbifolds}

The simplest case to consider is Type IIB superstring theory on
an orbifold.  For this, we begin with a spacetime manifold $X$,
and seek to divide by a finite group $G$ of symmetries 
of $X$.  $X$ is endowed with an orientation and spin structure, and
these are preserved by $G$.

$D$-brane configurations on $X/G$ are understood as $G$-invariant
configurations of $D$-branes on $X$ \mooredouglas. $G$ in general
may act in an arbitrary fashion on the gauge bundles supported
on the $D$-branes. A $D$-brane configuration,
as we have seen, represents in general a pair of bundles $(E,F)$.
This construction can be made in a completely $G$-invariant way
(see, for example \atiyah, section 2.3, for an introduction to such
matters), so we can assume that $G$ acts on $(E,F)$.  In tachyon condensation,
we should assume that a pair of bundles $(H,H)$ can be created or annihilated
only if $G$ acts on both copies of $H$ in the same way.  Otherwise,
the requisite tachyon field would not be $G$-invariant.

Pairs of bundles $(E,F)$ with $G$ action, 
modulo the relation $(E,F)=(E\oplus H,F\oplus H)$ for any bundle $H$
with $G$ action, form a group called $\K_G(X)$. ($\K_G(X)$ is called
the ``$G$-equivariant K-theory of $X$.'' See again \atiyah, section 2.3, 
for an introduction.)  We conclude that for Type IIB superstrings on $X/G$,
$D$-brane charge takes values in $\K_G(X)$.

For Type IIA, we similarly get $\K^1_G(X)$, and for Type I
we get $\KO_G(X)$.

The standard string theory formula for the Euler characteristic of an orbifold
$X/G$ (in Type II string theory)
has been shown \ref\stah{M. F. Atiyah and G. B. Segal, ``On
Equivariant Euler Characteristics,'' J. Geom. Phys. {\bf 6} (1989) 671.}
to coincide with the Euler characteristic in equivariant K-theory
(understood as the dimension of $\K_G(X)\otimes_\Z {\bf Q}$ minus
that of $\K^1_G(X)\otimes_\Z{\bf Q}$).  This is presumably related
to the fact that the Betti numbers of the orbifold, in the string theory
sense, determine the possible charges for Type IIB and Type IIA $p$-form
fields, and those charges actually take values in $\K_G(X)$ or $\K^1_G(X)$,
respectively.

\subsec{Involutions}

Now we specialize to the case that $G=\Z_2$, for which some additional
constructions are possible.  We denote the generator of $\Z_2$ as $\tau$.
Thus $\tau$ is a so-called ``involution'' of $X$, a symmetry with $\tau^2=1$.

Instead of simply dividing by the geometrical action of $\tau$ on
$X$, we have three additional options:

{\it (i)} We can divide by $\tau$ times $\Omega$, the operator that
reverses the orientation of a string.

{\it (ii)} We can divide by $\tau$ times $(-1)^{F_L}$, the operator
that acts as $-1$ or $+1$ on states in a left-moving Ramond or Neveu-Schwarz
sector.

{\it (iii)} We can divide by $\tau$ times the product $\Omega(-1)^{F_L}$.

More generally still, we could divide by  a finite symmetry group
$G$ of spacetime which has some elements
that act only geometrically and other elements that  act also via
 $\Omega$, $(-1)^{F_L}$, or $\Omega(-1)^{F_L}$.  For brevity, I will
not discuss this generalization, which combines the different cases.

I will briefly analyze the three types of $\Z_2$ action listed above.
Since $\Omega$ acts on 9-brane (or $\bar 9$-brane) bundles
by complex conjugation, in case {\it (i)}
 we want to consider
bundles $(E,F)$ that are mapped by $\tau$ to their complex conjugates.
\foot{If a bundle $E$ is defined with transition functions $g_{ij}$ relative
to an open cover $U_i$ of $X$, then the complex conjugate of $E$ is a
bundle $\bar E$ with transition functions $\bar g_{ij}$.}  Whenever we
say that a bundle, such as $E$, is mapped by $\tau$ to its complex conjugate
$\bar E$, 
we mean, to be more precise, that $\tau^*(E)$ - the pullback of $E$ by $\tau$
- is
isomorphic to $\bar E$, and that an isomorphism $\psi:\tau^*(E)\to
\bar E$ (obeying $(\psi\tau^*)^2=1$) is given.

Now we consider $(E,F)$ to be equivalent to $(E\oplus H,F\oplus H)$,
where $H$ is similarly mapped by $\tau$ to its complex conjugate.
\nref\atiyahtwo{M. F. Atiyah, ``K-Theory And Reality,'' Quart. J. Math.
Oxford (2) {\bf 17} (1966) 367, reprinted in {\it Michael Atiyah
Collected
Works}, op. cit.}
Such pairs make up a group that has been called ${\rm KR}(X)$
\atiyahtwo.
${\rm KR}(X)$ depends, of course, on the choice of $\tau$, but this
is not usually indicated explicitly in the notation.

Now let us try to interpret case {\it (ii)}.  Since $(-1)^{F_L}$ reverses
the sign of $D$-brane charge, $D$-brane configurations on $X/\Z_2$
should in this case be related to $D$-brane configurations  on $X$
whose K-theory class is {\it odd} under $\Z_2$.  This means that
$\tau$ maps the pair $(E,F)$ to $(F,E)$.  (This means that we
are given isomorphisms $\lambda:(E,F)\to (\tau^*F,\tau^*E)$
with $(\lambda\tau^*)^2=1$.)  We consider a trivial pair to be a pair
$(H,H)$ (with $H$ isomorphic with $\tau^*H$).  The group of such pairs
$(E,F)$ with $(E,F)$ equivalent to $(E\oplus H,F\oplus H)$ for any
such $H$, make a K-like group that does not seem to have been much
investigated mathematically.  The name $\K_\pm(X)$ has been proposed
for this group, and it has been argued by M. J. Hopkins that $\K_\pm(X)$
can be computed in terms of conventional equivariant K-theory as follows:
\eqn\hucc{\K_\pm(X)=\K_{\Z/2}^1(X\times \S^1).}
Here $\K_{\Z/2}$ is conventional equivariant cohomology for the group
$G=\Z_2$; $\Z_2$ acts on $X\times \S^1$ by the product of
the action of $\tau$ on $X$ and an orientation-reversing symmetry of $\S^1$.

Examples in which $\tau$ acts together with $(-1)^{F_L}$
 are interesting because only a few examples of stable nonsupersymmetric
 $D$-branes have been closely examined in the literature, and one of these
 \refs{\senone,\bergman} is of this type.  In those papers, an orbifold is
 considered in which space is
 $\R^9/\Z_2$, with $\Z_2$ acting by $-1$ on the last
 four coordinates of $\R^9$ and $+1$ on the first five, times $(-1)^{F_L}$.
 For this action of $\Z_2$ on $\R^9$, it has been shown by M. J. Hopkins
 (by using \hucc\ plus the periodicity theorem)
 that (for $\K_\pm$ with compact support) $\K_\pm(\R^9)=\Z$.
 Thus, we expect a stable $D$-brane configuration carrying an additive
 conserved charge.  This presumably is the configuration studied
 in \refs{\senone,\bergman}.  (In \senone, it was described as a tachyonic
 soliton on a brane-antibrane pair, and in \bergman\ 
 as a $D$-brane.)

\def\KR{{\rm KR}} 
The final case is type {\it (iii)}, in which $\tau$ acts via $\Omega(-1)^{F_L}$
and hence maps $(E,F)$ to $(\bar F,\bar E)$.  This combines KR theory
with $\K_\pm$.  The $D$-branes charges live in a group that might
be called $\KR_\pm(X)$. 

\subsec{Incorporation Of The $B$-Field}

So far in this paper, we have suppressed the role of the Neveu-Schwarz
$B$-field.  $B$ has a three-form field strength $H$, and a characteristic
class $[H]\in H^3(X,\Z)$.

When $[H]\not= 0$, it is no longer true that Type IIB $D$-brane charge
takes values in $\K(X)$.  Indeed, branes can be wrapped on a submanifold
$Z$ of spacetime only if (\witten, eqn. (3.13)) when restricted to $Z$
\eqn\kilp{[H]+W_3(Z)=0.}
($W_3(Z)$ is of order two, and so can be placed on the left or right
of this equation.)  For $[H]=0$, the condition is that $W_3(Z)=0$;
as we have seen in section 4, this is the right condition for K-theory.
For $[H]\not= 0$, the condition is clearly no longer the right one
for K-theory.

I will now argue that when $[H]$ is a torsion class (some examples of this
type were studied in \witten), $D$-brane charge takes values in a certain
twisted version of K-theory that will be described.  I do not know 
the right description  when $[H]$ is not torsion.

First recall the case with $[H]=0$.  We recall from section 4.3 that
the ``gauge bundle'' on a $D$-brane is twisted in a subtle but important
way.  Cover $X$ with open sets $U_i$, and describe $w_2(Z)$ by
a $\{\pm 1\}$-valued cocycle $\phi_{ijk}$ on $U_i\cap U_j\cap U_k$.
Then the gauge bundle of a $D$-brane is described by transition functions
$g_{ij}$ on $U_i\cap U_j$.  The transition functions for a vector
bundle would on $U_i\cap U_j\cap U_k$ obey $g_{ij}g_{jk}g_{ki}=1$.
Instead, in $D$-brane theory, the required condition is
\eqn\sons{g_{ij}g_{jk}g_{ki}=\phi_{ijk}.}
(In a footnote in section 3, we proved that this condition implies
that the $\phi_{ijk}$ obey the usual cocycle relation $\phi_{ijk}\phi_{ikl}=
\phi_{jkl}\phi_{ijl}$ on quadruple overlaps.)
For example, if $n=1$, functions $g_{ij}$ obeying \sons\ define
a ${\rm Spin}_c$ structure on $Z$.  This twisted condition 
was needed in section 4.3 to match to K-theory.

Since $[H]$ appears together with $W_3(Z)$ in the condition \kilp\ for
$D$-brane wrapping, one suspects that when $[H]\not= 0$,  a cocycle
defining $[H]$ should somehow be included in \sons.
I will describe how to do this when $[H]$ is torsion.
Consider the exact sequence
\eqn\ilpo{0\to \Z\underarrow{i}\R\to U(1)\to 0,}
with $i$ the inclusion of $\Z$ in $\R$.
The associated long exact sequence in cohomology reads in part
\eqn\cilop{\dots H^2(Z,\R)\underarrow{i}
H^2(Z,U(1))\to H^3(Z,\Z)\to H^3(Z,\R)\to \dots.}
We conclude that $[H]\in H^3(Z,\Z)$ maps to zero in $H^3(Z,\R)$
-- and so is a torsion class -- if and only if $[H]$ can be lifted
to an element in $H^2(Z,U(1))$, which we will call $H^*$.

The lift of $[H]$ to $H^*$, if it exists, is not necessarily unique.
Exactness of \cilop\ says that $H^*$ is unique modulo addition
of an element of the form $i(b)$, for any $b\in H^2(Z,\R)$.  
Suppose that $[H]$ is of order $n$ in $H^3(Z,\Z)$.  Then we can always
pick its lift to $H^2(Z,U(1))$ so that $H^*$ is of order $n$.
We cannot make the order of $H^*$ smaller than this, because
$mH^*=0$ implies $m[H]=0$.

Being of order $n$, $H^*$ can be represented by a cocycle 
valued in the $n^{th}$ roots of unity.  This means that on each
$U_i\cap U_j\cap U_k$, we are given an $n^{th}$ root of unity
$h_{ijk}$, obeying the cocycle relation on quadruple overlaps.

Now, since $[H]$ appears together with $W_3(Z)$ in \kilp, I propose
that the corresponding cocycles appear together in the generalization
of \sons.  The ``gauge bundle'' on a $D$-brane would thus be described
by transition functions obeying
\eqn\sonsi{g_{ij}g_{jk}g_{ki}=h_{ijk}\phi_{ijk}.}
It should be possible to check this directly via worldsheet global
anomalies.

Now let us specialize to the case of 9-branes (or $\bar 9$-branes).
In this case, we set $\phi_{ijk}=1$, since $X$ is spin.
Hence, 9-brane gauge bundles are described by transition functions
that obey
\eqn\bons{g_{ij}g_{jk}g_{ki}=h_{ijk}.}
The   direct sum of two such twisted bundles obeys obeys the same condition.
So it is possible to define a twisted K-group $\K_{[H]}(X)$ whose
elements are pairs $(E,F)$ of such twisted bundles subject to the usual
equivalence relation, which says that $(E,F)$ is equivalent
to $(E\oplus H,F\oplus H)$ for any $H$.  In \ref\donoka{P. Donovan
and M. Karoubi,  ``Graded Brauer Groups And K-Theory With Local
Coefficients,'' IHES Pub. {\bf 38} (1970) 5.}, it is shown
that \sonsi\ is the condition for a submanifold $W$ to determine
a class in this kind of K-theory.

\def\P{{\bf P}}
It is possible to describe twisted bundles more intrinsically, without
talking about open covers and transition functions (which have been
used here to try to keep things elementary).  This approach, which
is taken in the mathematical literature on $\K_{[H]}(E)$, proceeds
as follows.  If $E$ is a twisted
bundle, there are associated with it several ordinary bundles.  There is
a bundle $\P(E)$ of complex projective spaces.
The obstruction to deriving $\P(E)$ by projectivizing a vector bundle
is measured by the class $H^*\in H^2(Z,U(1))$.
Also, the endomorphisms of $E$ are valued in an ordinary vector bundle, whose
sections make an algebra $A(E)$.
In the mathematical literature, $\K_{[H]}(X)$ is defined in terms
of modules for the algebra $A(E)$.  One fundamental theorem 
\ref\groth{A. Grothendieck, ``Le Groupe de Brauer,'' section 1.6, in
{\it Dix Expos\'es Sur La Cohomologie Des Sch\'emas}, North
Holland (1968).} asserts that for any $[H]$ of finite order $n$,
there exists
a twisted bundle $E$ of some finite rank $m$ (which is always a multiple
of $n$).  Given this theorem, one proves as follows that $\K(X)$ and
$\K_{[H]}(X)$ are equivalent rationally.  Tensoring with $E$ gives
a map from $\K(X)$ to $\K_{[H]}(X)$; tensoring with $E^*$ (the dual
of $E$) gives a map back from $\K_{[H]}(X)$ to to $\K(X)$.  The composite
is multiplication by $m^2$ and so is an isomorphism rationally.

So far, our evidence that $D$-brane charge for Type IIB takes
values in $\K_{[H]}(X)$ is mainly formal: $\K_{[H]}(X)$  is a natural modified
version of $\K(X)$ that can be constructed from the data at hand,
and  extends \sons\ in a tempting way.  The analogous
statement for Type IIA is that $D$-brane charge is classifed by
$\K^1_{[H]}(X)$; for Type I, it should be classified by $\KO_{[H]}(X)$.

We will now give strong support for this picture by showing that
in the case of Type I, it is equivalent to something that is known
independently.  The worldsheet $\theta$ angles are odd under the projection
that reduces Type IIB to Type I, so must take the values $0$ or $\pi$ in
Type I.  This implies that for Type I, $[H]$ is of order 2, and the
cocycle $h_{ijk}$ takes values in $\{\pm 1\}$.  This means that
$H^*$ actually lies in the subgroup $H^2(X,\Z_2)$ of $H^2(X,U(1))$.

At $[H]=0$, for Type I the 9-branes and $\bar 9$-branes carry
$SO(n)$ vector bundles.  But if we turn on a non-zero $[H]$ which
is of order 2, then \bons\ becomes the condition for an $SO(n)/\Z_2$
bundle without vector structure, in the sense of
\ref\polchetal{M. Berkooz, R. G. Leigh, J. Polchinski, J. H. Schwarz,
N. Seiberg, and E. Witten, ``Anomalies, Dualities, and Topology Of
$D=6$ $N=1$ Superstring Vacua,'' Nucl. Phys. {\bf B475} (1996) 115,
hep-th/9605184.}.  
We recall that, just as the obstruction
to spin structure of an $SO(N)$ bundle $W$
is measured by a class $w_2(W)\in H^2(X,\Z_2)$, so the obstruction
to vector structure for an $SO(n)/\Z_2$ bundle $V$ is measured
by a class $\tilde w_2(V)\in H ^2(X,\Z_2)$.  $\tilde w_2(V)$ can be
represented by a $\{\pm 1\}$-valued cocycle $t_{ijk}$ on $U_i\cap U_j\cap
U_k$.  An $SO(n)/Z_2$ bundle whose vector structure is obstructed
by a cocycle $t$ has transition functions $g_{ij}$ that obey
$g_{ij}g_{jk}g_{ki}=t_{ijk}$.  
(The cocycle would cancel out if we take the matrices $g_{ij}$ in the
adjoint representation of $SO(n)$, or any other representation in
which the central element $-1$ of $SO(n)$ acts trivially.)
So \bons\ asserts that the cohomology class $H^*$ of the $B$-field
equals $\tilde w_2(V)$.
This statement is true \ref\sensethi{A. Sen and
S. Sethi, ``The Mirror Transform Of Type I Vacua In Six Dimensions,''
Nucl. Phys. {\bf B499} (1997) 45, hep-th/9703157.}.
Therefore, Type I $D$-brane charge takes values in $\KO_{[H]}(X)$,
giving strong encouragement to the expectation that the analogous
statements are true for Type II.  

\bigskip\noindent{\it Comparison To Cohomology Theory}

One point that now requires some discussion is why we can get this
kind of description only if $[H]$ is torsion.

K-theory is regarded as a generalized cohomology theory
(see for example section 2.4 of \atiyah).  To get
some intuition about K-theory with nonzero $[H]$, we might consider a hierarchy
in which at level one one has cohomology theory and     gauge fields,
and at level two one has K-theory and two-form $B$-fields.
(Level three might consist of elliptic cohomology and some stringy
construction, but that remains to be seen.) 

There is a notion of cohomology theory coupled to any gauge field
$A$ with zero curvature (cohomology with values in any flat bundle; the analogy
we are about to make is more precise if the bundle is a line bundle).
The level two analog  should be K-theory coupled
to any flat $B$-field.  A gauge field whose curvature is zero
is one whose Chern classes are torsion.  Similarly, a flat $B$-field
is one whose characteristic class $[H]$ is torsion.  A candidate
for K-theory coupled
to a flat $B$-field is our friend $\K_{[H]}(X)$.

What if $[H]$ is not torsion?  Let us compare to what happens for
cohomology.  One can couple differential forms to any vector bundle
with connection $A$, replacing the usual exterior derivative
by its gauge-covariant extension $d_A=d+A$.  But if the curvature of
$A$ is not zero, one no longer gets a cohomology theory, since $d_A^2\not= 0$.
By analogy, one cannot expect to define a generalized cohomology theory
when $[H]$ is not torsion; one must expect to go ``off shell'' in some way.
There is no obvious known mathematical theory; Type II string theory itself
may be the only candidate.  Perhaps there is an approach via noncommutative
geometry; so far, noncommutative geometry has been used to describe
$D$-branes coupled to flat but irrational and topologically trivial
$B$-fields \ref\connes{A. Connes, M. R. Douglas, and A. Schwarz, 
``Noncommutative Geometry And Matrix Theory: Compactification On Tori,''
JHEP {\bf 2} (1998) 3, hep-th/9711162.}.

\bigskip\noindent{\it Orbifolds With Discrete Torsion}

$D$-branes on an orbifold with discrete torsion 
 will lead, in view of the analysis in
 \ref\udoug{M. Douglas, ``$D$-Branes And Discrete Torsion,''
hep-th/9807235.},  to a mixture of two of the constructions
that we have considered.  In this case, instead of the K-theory of
pairs $(E,F)$ of bundles with $G$ action, one wants the K-theory
of such pairs with a projective action of $G$ (with a fixed cocycle
determined by the discrete torsion).
This presumably should be understood as a $G$-equivariant version of
$\K_{[H]}$.

\newsec{Stringy Constructions For Type I}

In this section, we will discuss the worldsheet construction of the zerobrane 
and $-1$-brane  of Type I.  
Actually, up to a certain point the discussion can be carried out
equally well for Type I or Type IIB.  However, since the generators
of $\KO(\S^9)$ and $\KO(\S^{10})$ are mapped to zero if one forgets
the reality condition of the bundles and maps KO-theory to ordinary K-theory,
we expect that the stable zerobrane and $-1$-brane of Type I correspond
to objects that are unstable if considered in Type IIB.  Thus
in the Type IIB description, we expect to see a tachyon that is removed
by the orientifold projection.

\subsec{The Zerobrane}

We consider first the zerobrane, which has already 
been analyzed \senthree; we will aim to clarify a few points.
(The discussion applies equally well to the eightbrane, as we briefly note
later.)
From the point of view of K-theory, the Type I zerobrane is described
by the same tachyon field $T=\vec\Gamma\cdot \vec x$ 
as the supersymmetric
branes.  This suggests that we should interpret the zerobrane as a $D$-brane,
much like the more familiar supersymmetric branes.

The naive idea is then to introduce in Type I a $D$-particle -- 
located, say, at $x^1=\dots = x^9 = 0$.
One immediately runs into the following oddity
(which corresponds to the factor of $\sqrt 2$ in the multiplicity
of states noted at the end of section 4).
Type I superstring theory also has  9-branes, so there are $0$-9 open
strings that must be quantized. In the Neveu-Schwarz 0-9 sector,
the fermions $\psi_1,\dots,\psi_9$ (superpartners of $x^1,\dots,x^9$)
have zero modes which we may call $w_1,\dots,w_9$.  We therefore
have to quantize an {\it odd}-dimensional Clifford algebra,
\eqn\humboo{\{w_i,w_j\}=2\delta_{ij},\,\,i,j=1,\dots,9.}
There is, however, no satisfactory quantization of such an odd-dimensional
Clifford algebra.\foot{In the Ramond 0-9 sector, one gets the
same basic problem of an odd-dimensional Clifford algebra -- in this
case $\psi_1,\dots,\psi_9$ have no zero mode but $\psi_0$ does.}
The nine-dimensional Clifford algebra has two irreducible representations,
each of dimension 16, and differing simply by $w_i\to -w_i$.
(The two representations of the Clifford algebra are equivalent as
representations of ${\rm Spin}(9)$.)  In one representation, the product
$\bar w= w_1w_2\cdot \dots \cdot w_9$, which is in the center of the
Clifford algebra, is represented by $+1$, 
and in the other representation,
$\bar w=-1$.   Generally, in quantum field theory,
we should use an irreducible representation of the algebra of observables.
In this case, we have the problem that in an irreducible representation,
there is no operator $(-1)^F$ that anticommutes with the $w_i$ (such
an operator would clearly change the sign of $\bar w$ and so interchange
the two representations of the Clifford algebra).
Without a $(-1)^F$ operator,
we cannot make sense of the worldsheet sum over spin structures.
To have a $(-1)^F$ operator, we must include both representations of the 
Clifford algebra (and let $(-1)^F$ exchange them).  But what is a natural
explanation of this doubling of the worldsheet spectrum?

To account for this doubling, we should have a tenth fermion zero mode, say 
$\eta$, on the $0$-9 string.  Then the doubling of the spectrum arises because
the (unique) irreducible representation  of the ten-dimensional Clifford 
algebra
 is 32-dimensional, and decomposes
under $w_1,\dots,w_9$ as the direct sum of the two 16-dimensional 
representations of the nine-dimensional Clifford algebra. 
The operator that anticommutes with $\bar w$ is $\eta$, and the
operator that acts as $(-1)^F$ on the zero modes (anticommuting with
$\eta$ as well as the $w$'s) is $\eta\bar w$.  Making the GSO projection
has the effect, on the string ground state, of reducing from 32 states
to a single irreducible 16-dimensional representation of the smaller Clifford
algebra.  Thus, for the sake of counting states, the spectrum is the same
as if we use an irreducible representation of the original nine-dimensional
Clifford algebra and do not make a GSO projection.   But adding the extra
fermion zero mode and making the GSO projection gives a way to get this
spectrum that is more coherent with the rest of string theory.

To obtain this extra fermion zero mode for 0-9 strings, 
we postulate that on any boundary
of an open string that lies on the zerobrane, there is a field
$\eta(\tau)$ ($\tau$ being a parameter along the boundary)
with Lagrangian
\eqn\kson{L=i\int d\tau \,\,\eta{d\eta \over d\tau}.}
Quantization of this Lagrangian gives the required $\eta$ zero mode.

Now we can describe some rules for worldsheet computations.
Consider a worldsheet $\Sigma$ with a boundary component $S$ (which is
a circle, of course) on the zerobrane.
The spin structure of $\Sigma$
when restricted to $S$ may be in either the Neveu-Schwarz or Ramond
sector.  In the NS sector, $\eta$ is an antiperiodic function on $S$, and
in the Ramond sector, it is periodic.
We assume that the worldsheet path integral includes an integral over $\eta$
which (if any vertex operators on $S$ are independent of $\eta$) 
is simply
\eqn\jcza{\int D\eta(\tau)\exp\left(i\oint_S d\tau \eta{d\eta\over d\tau}
\right).}
This integral is easy to calculate.  It equals 0 in the Ramond sector,
and $\sqrt 2$ in the Neveu-Schwarz sector.  The vanishing in the Ramond
sector arises because
there is an $\eta$ zero mode in the path integral -- the
constant mode of $\eta(\tau)$.  As for the NS path integral, as the Hamiltonian
is zero, one would expect it to count the number of states obtained
by quantizing the field $\eta$.  We cannot quite give this path integral
that interpretation, because this system has no natural quantization;
trying to quantize it, we get a one-dimensional Clifford algebra
(generated by $\eta$), which like any odd-dimensional Clifford algebra
has no natural quantization.  However, if we had {\it two} $\eta$ fields,
the path integral
\eqn\jikks{\int D\eta_1(\tau)\,D\eta_2(\tau)\,\,\exp\left(
i\sum_{i=1}^2\oint_S\eta_i{d\eta_i\over d\tau}\right) }
would equal 2, because in this case the quantum system (a two-dimensional
Clifford algebra) can be naturally quantized and has two states.
So for one $\eta$ field, the path integral equals $\sqrt 2$.   All the usual
factors in the worldsheet path integral must be supplemented with this
factor, which   was found in \senthree\ from another point of view.
Because of this factor, a zerobrane of this kind in Type IIB has a mass greater
than the mass of a conventional Type IIA zerobrane (with the same
values of $\alpha'$ and the string coupling) by a factor of $\sqrt 2$.

\def\O{{\cal O}}
The vanishing of the $\eta$ path integral in the Ramond sector
assumes that the 0-0 vertex operators, inserted on the zerobrane
boundary, are independent of $\eta$.  What in fact do those operators
look like?  Since $d\eta/d\tau=0$ by the $\eta$ equation of motion,
and $\eta^2=0$ by fermi statistics, the possible vertex operators are
at most linear in $\eta$, and hence take the form $\O(X,\psi)$
or $\eta\O'(X,\psi)$.  ($X$ and $\psi$ are the worldsheet matter fields,
and for brevity we omit ghosts from the notation.)
A simple generalization of the reasoning by which we analyzed the
path integral \jcza\ shows that an amplitude with an odd number of
vertex operator insertions of the form $\eta\O'$ on $S$ receives
a contribution only from the Ramond sector (that is, from spin structures
that restrict on $S$ to the Ramond sector), while an amplitude with
an even number of such insertions receives a contribution only
from the NS sector.

Note that $\eta$ is of conformal
dimension zero, so for $\eta \O'$ to be of dimension one, $\O'$ is
of dimension one.
The GSO projection as usual projects out the NS tachyon of type $\O$.
But since $\eta$ is odd under $(-1)^F$,  the states with vertex
operators of type $\eta\O'(X,\psi)$ undergo
 the opposite of the usual GSO projection.  The $\eta \O'$ tachyon therefore 
survives
the GSO projection.

Thus, this zerobrane, regarded as an excitation of Type IIB superstring
theory, is tachyonic.  To understand what happens for Type I,
one must still analyze the $\Omega$ projection,
which reverses the orientation of the open string.  The zerobrane will be 
stable
in Type I if the $\Omega$ projection removes the tachyon from the 0-0 sector.

That it does so has been deduced by Sen from another construction
(in which the zerobrane is built by a marginal deformation of
a $1$-$\bar 1$ system \senthree).  Sen also suggested the following
direct approach to the question.  Since we know how the $\Omega$
operator acts on closed strings, we can deduce how it acts on open
strings by looking at transitions between 0-0 open strings and closed
strings.  The simplest worldsheet describing such a transition is
a disc $D$, with  boundary ending on the zerobrane.  We consider an
amplitude with one  0-0 vertex operator on the boundary of the disc,
and one closed string operator in the interior.
We consider the case that the states making the transition are bosonic.
We want to look at a two-point function of the form
$\langle {\cal W}\cdot \eta\O'\rangle$, with ${\cal W}$ a closed
string vertex operator in the interior of the disc, and $\eta\O'$ an
open string vertex operator.

As we have discussed before, since there are an odd number of $\eta\O'$
insertions, such an amplitude can be nonzero only if the spin structure
on the boundary of $D$ is in the Ramond sector.  This is possible only
if the vertex operator ${\cal W}$ is a Ramond-Ramond vertex operator,
which creates a ``cut'' in the worldsheet fermions.
 In fact, we take ${\cal W}$ to
be the vertex operator of the massless RR scalar of Type IIB.
We can write this in the $(-1/2,-1/2)$ picture as
${\cal W}=e^{-(\phi(z)+\tilde\phi(\bar z))/2}
k\cdot \Gamma_{\alpha\beta}S^\alpha(z)\tilde S^\beta(\bar z)e^{ik\cdot X}$
with $z$ a complex parameter on the disc, $k$ the momentum,
and $S$ and $\tilde S$ the left and right-moving spin fields.
This particular state is odd under $\Omega$ and is projected out in
reducing to Type I.  So we can show that the 0-0 tachyon is odd under
$\Omega$ by showing that it can make a transition to the RR scalar.

The vertex operator of the 0-0 tachyon, in the $-1$ picture,
is $\eta{\cal O}'(\tau)=\eta e^{-\phi}e^{iq\cdot X}$.  ($q$
points  in the ``time'' direction, since the 0-0 tachyon propagates
only on the zerobrane.)
The matrix element $\langle {\cal W}\cdot\eta{\cal O}'\rangle$ is nonzero,
since $\eta$ gets an expectation value due to the zero mode, while nonvanishing
of the matter and ghost matrix element is equivalent to the statement
\polch\ that Type IIA zerobranes carry RR charge.
 
\nref\gimon{E. Gimon and J. Polchinski, ``Consistency Conditions For
Orientifolds And $D$-Manifolds,'' Phys. Rev. {\bf D54} (1996) 1667,
hep-th/9601038.}
One could in exactly the same way construct a Type I eightbrane; $\eta$
is included in the same way.
Suppose that we want instead a Type I $p$-brane for $p=2,4$, or 6.
For $p=2,6$, we run into the following.  After adding the $\eta$ field,
we have $p+2$ fermion zero modes in the NS sector of the $p$-9 string.
This is an even number, which enables us to make sense of the GSO projection.
But after imposing the GSO projection, we are left with the chiral spinors
of $SO(1,p+1)$ (obtained by quantizing $p+2$ fermion zero modes, one of
which has ``timelike'' metric).  These chiral spinors are complex, 
contradicting
the fact that the $p$-9 wavefunctions should be real. So there is apparently
no twobrane or sixbrane.  Likewise, for a fourbrane, we would meet chiral
spinors of $SO(1,5)$, which are pseudoreal, rather than real.
So there is no fourbrane either.\foot{If we include symplectic fourbrane
Chan-Paton factors to make the 4-9 strings real, we find that the
4-4 spectrum has a tachyon.}  These results agree with $\KO(\S^3)
=\KO(\S^5)=\KO(\S^7)=0$, $\KO(\S^1)=\KO(\S^9)=\Z_2$.

\def\KSp{{\rm KSp}}
Suppose, however, that the ninebranes were quantized with symplectic
rather than orthogonal Chan-Paton factors.  (There is no supersymmetric
way to do this with tadpole cancellation, but up to a certain point
we can consider such a theory anyway.)  $D$-brane charge would then
take values in a K-group $\KSp(X)$ 
whose elements are pairs $(E,F)$ of symplectic
bundles, modulo the usual sort of equivalence relation.  
The symplectic or pseudoreal Chan-Paton
factors of the ninebrane would cancel the reality problem for the fourbrane
while creating one for the zerobrane and eightbrane.  So this kind
of theory has a fourbrane but no other even-dimensional branes.
This is in agreement with the results of Bott periodicity,
according to which $\KSp(\S^5)=\Z_2$, $\KSp(\S^1)=\KSp(\S^3)=\KSp(\S^7)
=\KSp(\S^9)=0$.

\bigskip\noindent{\it Comparison To Gimon-Polchinski}

One might wonder whether we instead could use the arguments of Gimon and
Polchinski \gimon\ to learn how $\Omega$ acts on the 0-0 strings.
Their approach would entail examining the operator product of
0-9 and 9-0 vertex operators.  These vertex operators require
some novelty, since spin fields for an odd number of fermions are
not usually considered in conformal field theory, and the part of the
0-9 vertex operator involving $\eta$ is also somewhat unusual.
I will not try to make this analysis here.

\bigskip\noindent{\it Spinor Quantum Numbers And Multiplicative Conservation 
Law}

To tie up the discussion with what was said in section two,
we note finally that as explained in \senthree, the Type I zerobrane
transforms in the spinor representation of $SO(32)$. 
In fact, in the Ramond 0-9 sector, there are two fermion zero modes
(modes of $\eta$ and $\psi^0$), whose quantization gives two states,
of which one obeys the GSO projection.  Allowing for the 9-brane Chan-Paton
factors, this gives a single $SO(32)$ vector of fermion zero modes,
whose quantization gives a spinor representation of $SO(32)$.  (There also
are fermion zero modes, coming from other sectors,
 that are $SO(32)$-invariant.)

Suppose we are given a set of $k$ coincident Type I zerobranes.
Then the tachyon vertex operators depend on a $k\times k$ matrix
$M$ which acts on the Chan-Paton factors, and takes the form $\V(M)= 
M\cdot \V_0$, where $\V_0$ is the tachyon vertex operator as analyzed above.
The $\Omega$ projection, in addition to the action found
above, maps $M\to M^T$.  If $M$ is antisymmetric, it is odd under $\Omega$,
so the $\eta\O'$ tachyons with antisymmetric $M$ survive the $\Omega$ 
projection.
  An antisymmetric $M$ has an even
number of nonzero eigenvalues.  Every pair of nonzero eigenvalues describes
(presumably) the flow toward annihilation of a pair of zerobranes.
So the zerobrane number is conserved only modulo 2, as expected.

\subsec{The $-1$-Brane}  

We have seen in section 4.5 that the Type I $-1$-brane is understood
in K-theory as a Type IIB $-1$ brane-antibrane pair that are exchanged
by complex conjugation.  So, in a worldsheet construction, we will try
to understand this object by starting in Type IIB
with a $-1$-brane and antibrane, and assuming that they
are exchanged by worldsheet orientation reversal $\Omega$.

Somewhat more generally, 
consider in Type IIB a system consisting of a coincident
$p$-brane and $\bar p$-brane, for $p=-1, 3, $ or $7$.  These values
are selected because they are the values for which $\Omega$  reverses
the sign of RR charge and maps Type IIB $p$-branes to $\bar p$-branes.
As we reviewed in section three, the $p$-$\bar p$ system in Type IIB
has a tachyon which arises because the GSO projection for $p$-$\bar p$
open strings is opposite to the usual GSO projection.  The $p$-$\bar p$
system hence exists for Type IIB but is unstable.

The only hope of stabilizing it for Type I is that the $\Omega$
projection might remove the tachyon from the $p$-$\bar p$ system.
Note that there would be no hope of this for $p=1,5,$ or 9 -- the values
for which RR charge is $\Omega$-invariant.  In these cases, $\Omega$ maps
$p$-branes to $p$-branes and $\bar p$ to $\bar p$; so it maps
$p$-$\bar p$ open strings to $\bar p$-$p$ open strings.  $\Omega$
hence cannot eliminate the open string tachyon for these values of $p$;
it merely relates the $p$-$\bar p$ tachyon to the $\bar p$-$p$ tachyon.

Instead, for $p=-1,3,7$, $\Omega$ exchanges $p$-branes with $\bar p$-branes.
Hence, for those values of $p$, $\Omega$ maps $p$-$\bar p$ open strings
to themselves (as a result of exchanging the two ends of the string
but also turning $p$ into $\bar p$ and vice-versa), and likewise
maps $\bar p$-$p$ open strings to themselves.  Hence it is conceivable
that for $p=-1,3,$ or $7$, the $\Omega$ projection might remove the tachyon
and stabilize the $p$-$\bar p$ system.  

For $p=1,5$, or 9, $\Omega$ maps $p$-$p$ open strings to themselves
and hence, if one considers $N$ parallel $p$-branes, $\Omega$ reduces
the gauge group from $U(N)$ to $SO(N)$ or $Sp(N)$.  Which reduction is made
 depends on how $\Omega$ acts.  Gimon and Polchinski \gimon\ gave
a systematic procedure for showing that one gets $SO(N)$ for $p=1$ or 9
and $Sp(N)$ for $p=5$.

For $p=-1,3$ or 7, there is no analogous question of reduction of the
gauge group.  Given a system of $N$ parallel $p$-brane-antibrane pairs,
$\Omega$ maps the $p$-$p$ open strings to $\bar p$-$\bar p$ open strings
and hence identifies the $U(N)$ of the $p$-$p$ sector with the $U(N)$
of the $\bar p$-$\bar p$ sector; the unbroken gauge group is a diagonal
$U(N)$ regardless of any phases in the action of $\Omega$.

In summary then:

(1) For $p=1,5$, or $9$, one must analyze the $\Omega$ action on the $p$-$p$
sector to understand
what kind of gauge group the branes carry, but there is no sharp question
about the $p$-$\bar p$ sector, which is simply mapped to $\bar p$-$p$.

(2) For $p=-1,3$, or $7$, one must analyze the $\Omega$ action on
the $p$-$\bar p$ sector to determine whether the $p$-$\bar p$ configuration
is stable in Type I, but there is no sharp question about the $p$-$p$ sector,
which is simply mapped to $\bar p$-$\bar p$.

\def\V{{\cal V}}
Though the questions of interest are thus rather different for
$p=-1$, 3, or 7 than what they are for $p=1,5$, or 9, they can be answered
in the same way, using arguments by Gimon and Polchinski \gimon.
The basic idea is to let $\V$ be a physical $p$-9 vertex operator.
Denote as $\V^\Omega$ the transform of $\V$ by $\Omega$; it is
a $9$-$p$ or $9$-$\bar p$ vertex operator depending on whether $p$
is congruent to $1$ or $-1$ modulo 4.  A vertex operator that arises
as a pole in the $\V\cdot \V^\Omega$ operator product would be a physical
$p$-$p$ or $p$-$\bar p$ vertex operator; by identifying it, we can learn
which $p$-$p$ or $p$-$\bar p$ vertex operators survive the $\Omega$ projection.

In the $p$-$9$ NS sector, there are $9-p$ fermion zero modes.
In fact, the worldsheet fermions, which we label as
$\psi_i,\,i=0,\dots, 9$ (with the $p$-brane spanning the first $p+1$
coordinates) have zero modes precisely if $i>p$.  The NS ground
state thus transforms in a spinor representation of $SO(9-p)$. 
(For our present purposes, the number $9-p$ is even, an important fact, 
as we have seen!)
After imposing the GSO projection, the 9-$p$ ground states actually
transform as an $SO(9-p)$ spinor of definite chirality, say positive.
The vertex operator of such a state, in the $-1$ picture, is of the
form $\V(\epsilon)=e^{-\phi}\epsilon^\alpha S_\alpha e^{ik\cdot X}$, where 
$S_\alpha$ are the positive chirality spin fields and $\epsilon$ is
a $c$-number wave-function in the spinor representation.

  It is convenient to
divide the worldsheet fermions in the $p$-$\bar p$ sector as
$\psi_i',\,i=0,\dots,p$ and $\psi_j'',j>p$.  Neither $\psi_i'$ nor
$\psi_j''$ has a zero mode in the NS sector.  
Consider a $p$-$\bar p$ vertex operator
of the general kind $\O(X,\psi',\psi'')$.  The vertex operators also may
contain derivatives of the fields with respect to $\tau$ (a parameter on
the boundary of the worldsheet), but this is not shown in the notation.
To determine how a vertex operator transforms
under $\Omega$, one includes a factor of $-1$ for each $\tau$ derivative,
a factor of $-i$ for each $\psi'$, and a factor of $+i$ for each $\psi''$.
(Note here that $\psi'$ and $\psi''$ obey opposite boundary conditions
on the boundary of the worldsheet and so transform under $\Omega$ with
opposite phases.  The phases are $\pm i$, not $\pm 1$, because 
$\Omega^2=(-1)^F$
for open strings, and $\psi'$, $\psi''$ are both odd under $(-1)^F$.)
In addition to these factors, $\Omega$ acting on the $p$-$\bar p$ sector
gives a fixed additional phase $\alpha$, coming from its action on the
Chan-Paton wavefunction. This phase can be determined by looking at the 
$\Omega$ 
action on $\V(\epsilon)\cdot \V(\epsilon)^\Omega$; $\alpha$ must be such that
that state, which we know must be present in the spectrum, survives
the $\Omega$ projection.  

In particular, for $p=-1,3,$ or 7, we want to determine
whether $\V(\epsilon)\cdot \V(\epsilon)^\Omega$
transforms the same way as the $p$-$\bar p$ tachyon or oppositely to it.
This will determine whether the tachyon is present in the spectrum or not.
Note that the vertex operator for the tachyon is in the zero picture
${\cal W}=
k\cdot \psi' \,e^{ik\cdot X}$.  It transforms under $\Omega$ as $-i\alpha$.

As explained by Gimon and Polchinski, $\V(\epsilon)\cdot \V(\epsilon)^\Omega$ 
transforms under $\Omega$ as $i^{(9-p)/2}\alpha$.  The factor of
$i^{(9-p)/2}$ arises because if $\epsilon$ is a highest weight state,
then $V(\epsilon)\cdot V(\epsilon)^\Omega\sim (\psi'')^{(9-p)/2}$.
Since this state must survive in the spectrum, one has $\alpha=-i^{(9-p)/2}$.
Hence the $p$-$\bar p$ tachyon transforms as $i^{1+(9-p)/2}$.
It is therefore projected out for $p=-1$ or 7, but survives for $p=3$.
Hence, Type I has a stable $-1$-brane and a stable sevenbrane, but no
stable threebrane.  This result reflects the facts $\KO(\S^2)=\KO(\S^{10})
=\Z_2$, $\KO(\S^6)=0$.

\def\KSp{{\rm KSp}}
It is interesting to note that this result would be reversed if the
ninebranes are quantized with symplectic rather than orthogonal
Chan-Paton factors.  Then $\V(\epsilon)$ carries a symplectic
Chan-Paton label which gives an extra minus sign in the transformation
of $\V\cdot \V^\Omega$.\foot{If one endows $\V$ with a symplectic index
and tries to find in $\V\cdot \V^\Omega$ the same $p$-$\bar p$ operator
as before, one instead gets zero because of antisymmetry in the contraction
of the symplectic indices of the operators.  To get a nonzero result,
one may, for example,
 look in the operator product $\V(-\tau)\V^\Omega(\tau)$ for
an operator that appears with a coefficient odd under $\tau\to -\tau$;
this oddness corresponds to an extra factor of $-1$ in the transformation
under $\Omega$.}
  As a result, in such a theory, there is a stable
threebrane but no stable $-1$-brane or sevenbrane.  This result reflects
the facts that $\KSp(\S^6)=\Z_2$, $\KSp(\S^2)=\KSp(\S^{10})=0$,
so in such a theory, one expects a threebrane but no $-1$-brane or sevenbrane.

Finally, note that Gimon and Polchinski prove \gimon\ using the
above factor of $i^{(9-p)/2}$ that for a system of only $p$-branes
with $p=-1,3,$ or 7, one cannot define an $\Omega$ projection.
This argument uses the fact that for $p$-branes only, $\Omega(\Omega^T)^{-1}$
should act on the Chan-Paton labels as a $c$-number.  If one has both
$p$-branes and $\bar p$-branes, then $(-1)^F$ acts nontrivially on the
Chan-Paton labels, so the $\Omega$ action is more involved, and the
problem found in \gimon\ does not arise.

\bigskip\noindent{\it Symmetry Breaking And $\Z_2$ Conservation Law}

To compare to what was said in section 2, let us now show that the
amplitude due to a Type I $-1$-brane is odd under the disconnected
component of $O(32)$.  As in section 2, this happens because in the
presence of a $-1$-brane, there is a single $SO(32)$ vector of fermion
zero modes (plus other zero modes that are $SO(32)$-invariant).
They arise from the  $-1$-9 Ramond strings.
In quantizing this sector, there are no worldsheet fermion zero modes, so
there is a unique 
ground state for each value of the Chan-Paton labels; by worldsheet
supersymmetry its energy is zero.  For one of the
two possible $-1$-brane labels, this state survives the GSO projection;
it also transforms as a vector of $SO(32)$ because of the 9-brane Chan-Paton
labels, and this gives the expected multiplet of zero modes.

That the number of such $-1$-branes is conserved only modulo 2 may be seen,
as for zerobranes, by observing that in a system of $k$ coincident
Type I $-1$-branes, there is a tachyon described by a $k\times k$
antisymmetric matrix $M$.

\bigskip

This work was supported in part by NSF Grant PHY-9513835.
I would like to thank P. Deligne, D. Freed,
M. J. Hopkins,  S. Martin,
G. B. Segal, and A. Sen for many helpful explanations.

\listrefs
\end